\documentclass[11pt]{article}

\usepackage[latin1]{inputenc}
\usepackage{graphicx}
\usepackage{color}
\usepackage{graphicx}
\usepackage{comment}
\usepackage[mathscr]{euscript}
\usepackage{algorithm}
\usepackage[noend]{algpseudocode}
\usepackage{enumitem}
\usepackage{lscape}
\usepackage{makecell}
\usepackage{lscape}
\usepackage{verbatim}
\usepackage[colorlinks=true, linkcolor=blue, citecolor=blue, urlcolor=blue]{hyperref}
\usepackage{booktabs, multirow}
\usepackage{amssymb,amsmath}
\usepackage{float}
\usepackage{xcolor}
\usepackage[round]{natbib}

\def\0{\bf 0}
\def\1{\bf 1}
\def\B{\bf B}
\def\I{\bf I}
\def\PP{\bf P}
\def\RR{\bf R}
\def\r{\bf r}
\def\s{\bf s}
\def\u{\bf u}

\def\y{\bf y}

\def\bbeta{\mbox{\boldmath{$\beta$}}}
\def\ttheta{\mbox{\boldmath{$\theta$}}}
\def\pphi{\mbox{\boldmath{$\phi$}}}
\def\ggamma{\mbox{\boldmath{$\gamma$}}}
\def\ddelta{\mbox{\boldmath{$\delta$}}}
\def\SSigma{\mbox{\boldmath{$\Sigma$}}}

\begin{document}


\title{\textbf{Automatic cross-validation in structured models:\\ Is it time to leave out leave-one-out?}}

\author{Adin, A.$^{1,2}$, Krainski, E.$^{3}$, Lenzi, A.$^{4}$, Liu, Z.$^{5}$, Mart{\'i}nez-Minaya, J.$^{6}$, and Rue, H.$^{3}$\\
    \small {\textit{$^1$ Department of Statistics, Computer Sciences and Mathematics, Public University of Navarre, Spain.}} \\
    \small {\textit{$^2$ Institute for Advanced Materials and Mathematics, InaMat$^2$, Public University of Navarre, Spain.}}\\
    \small {\textit{$^3$ Statistics Program, Computer, Electrical and Mathematical Sciences and Engineering Division,}} \\
    \small {\textit{King Abdullah University of Science and Technology (KAUST).}}\\
    \small {\textit{$^4$ School of Mathematics, University of Edinburgh, Scotland.}}\\
    \small {\textit{$^5$ RIKEN Center for AI Project, Tokyo, Japan.}}\\
    \small {\textit{$^6$ Department of Applied Statistics, Operations Research and Quality, Universitat Polit{\`e}cnica de Val{\`e}ncia, Spain.}}\\
} \date{}

\makeatletter \pdfbookmark[0]{\@title}{title} \makeatother

\maketitle

\begin{abstract}
Standard techniques such as leave-one-out cross-validation (LOOCV) might not be suitable for evaluating the predictive performance of models incorporating structured random effects. In such cases, the correlation between the training and test sets could have a notable impact on the model's prediction error. To overcome this issue, an automatic group construction procedure for leave-group-out cross validation (LGOCV) has recently emerged as a valuable tool for enhancing predictive performance measurement in structured models.
The purpose of this paper is (i) to compare LOOCV and LGOCV within structured models, emphasizing model selection and predictive performance, and (ii) to provide real data applications in spatial statistics using complex structured models fitted with INLA, showcasing the utility of the automatic LGOCV method.
First, we briefly review the key aspects of the recently proposed LGOCV method for automatic group construction in latent Gaussian models. We also demonstrate the effectiveness of this method for selecting the model with the highest predictive performance by simulating extrapolation tasks in both temporal and spatial data analyses. Finally, we provide insights into the effectiveness of the LGOCV method in modelling complex structured data, encompassing spatio-temporal multivariate count data, spatial compositional data, and spatio-temporal geospatial data.
\end{abstract}


\bigskip

\section{Introduction} \label{sec:Intro}

Cross-validation is a general approach used to evaluate the predictive performance of a statistical model in the absence of new data, and it is commonly used to compute score rules for model comparison or selection. The fundamental concept underlying cross-validation is to split the observed data into a \textit{training set} (the sample data used for estimating the model's parameters) and a \textit{testing set} (the set of data points used to compute prediction errors based on the trained model) multiple times and use the combined prediction error to estimate the predictive accuracy of a model \citep{gelman1995bayesian,hastie2009elements}.

The $k$-fold cross-validation (KCV) is a widely used variant of cross-validation, where data is first randomly partitioned into $k$
equally (or nearly equally) sized mutually exclusive groups or folds, so that one fold is used to test the performance of the model fitted on the remaining $k-1$ folds. The procedure is repeated $k$ times, with each repetition employing a distinct subset of the observed data as the validation set. However, a poorly chosen value for $k$ may lead to a misrepresentative idea of the predictive performance of the model. For instance, it could result in a score with a high variance if the data used to fit the model give rise to highly variable test error measures. Alternatively, it could lead to a high bias, such as an overestimate of the performance of the model. Although no formal rule exists, a value of $k$ equal to 5 or 10 is widely prevalent in the fields of statistics and applied machine learning as a trade-off between computational burden and bias reduction of the predictive error \citep{hastie2009elements,kuhn2013applied}.

Leave-one-out cross-validation (LOOCV) represents limiting case of KCV, wherein individual observations are systematically excluded, one at a time, to test the model's performance trained on the remaining data. That is, the data is divided into $k=n$ folds (where $n$ is the size of the observed data) so that each data point is sequentially used as a testing set. The main advantages of LOOCV are that (i) the size of the training set closely approximates the full data set and, (ii) it provides an approximately unbiased estimate for the predictive error.
While the exact computation of LOOCV is a ``simple'' and straightforward strategy, as we only need to fit models on all possible training sets and compute the utility function on the corresponding testing set, this method can become practically unfeasible due to the prohibitive computational cost of fitting the model as many times as the number of observations in the data. For this reason, several methods have been proposed to efficiently compute fast approximations of LOOCV. See, e.g. \cite{burkner2021efficient,Heldetal2010cpo,Liu2022} and references therein.

Standard cross-validation techniques like KCV and LOOCV operate under the assumption that data is sampled independently from a joint distribution, so that it can be freely split into training sets that represent the observed data and testing sets that represent the unobserved data \citep{arlot2010survey}. However, it is well-known that these techniques might not be suitable to measure the predictive
performance of a model for correlated data, where commonly hierarchical models with structured random effects are considered to
capture the temporal, spatial and/or hierarchical clustering dependence structures within the data \citep{roberts2017cross,rabinowicz2022cross}. In such situations, the correlation existing between the training and testing sets can have a notable impact on the model's prediction error, and consequently affect the decision-making process for model selection, since standard cross-validation techniques are generally over-optimistic about the predictive performance of structured models.

A common approach for enhancing independence in cross-validation techniques is to adapt LOOCV to structured models by adjusting the
training sets for each testing set, a process frequently determined by the specific demands of prediction task. For example, in the context of time series data (or other intrinsically ordered data) the prediction task could encompass various objectives such as interpolating a missing value in the observed data, predicting the response in an independent replicate of the observed time series, or forecasting the response in future time points. For the first case, using LOOCV could be a suitable strategy because assessing the model's predictive performance by removing a single data point replicates the interpolation task effectively. However, a different cross-validation scheme should be used to evaluate the model's ability when a time series model is used for forecasting purposes (extrapolation task). A detailed comparison of different cross-validation approaches for time
series forecasting is described in \cite{bergmeir2012use}.

Therefore, when handling structured models, it is essential to adapt the cross-validation design to ensure the calculation of dependable scoring rules for model comparison or selection. In practice, many real data analyses demand the use of complex models with multiple correlation structures such as random effects at different aggregation levels, making it very difficult to determine the optimal design and adapting it to each of the testing points. These problems are exacerbated when dealing with complex structures on large data sets. To address both issues concurrently, \cite{Liu2022} introduced an automatic group construction procedure for leave-group-out cross-validation (LGOCV) to estimate the predictive performance of structured models by deriving an efficient and accurate approximation of LGOCV for latent Gaussian models within the INLA framework \citep{rue2009approximate,van2023new}.

The aim of the current paper is two-fold: (i) to compare the behavior of predictive measures based on leave-one-out and leave-group-out
cross-validation techniques within temporally or spatially structured models, and (ii) to provide real data applications of complex structured models fitted with INLA in the context of spatial statistics, while demonstrating the use of the automatic LGOCV method. In addition, we conduct a simulation study to evaluate the performance of the automatic group construction method for model comparison (see Supplementary Material).

The rest of the paper is structured as follows. Section~\ref{sec:LGOCV} briefly describes the most relevant aspects of the LGOCV method for latent Gaussian models. In Section~\ref{sec:Validation}, we compare the extrapolation performance of several structured models using both temporal and spatial data. Our objective is to emphasize disparities between model selection driven by conventional criteria for Bayesian model comparison and predictive performance measures based on LOOCV and the recently proposed automatic LGOCV method. We also validate the models' predictive performance by imitating real extrapolation tasks. Section~\ref{sec:Applications} presents three different real data applications in the field of spatial statistics.
Specifically, a joint modelling of pancreatic cancer mortality and incidence data in England using spatio-temporal multivariate disease mapping models (Section~\ref{sec:DiseaseMapping}), the estimation of the membership proportions to the different genetic clusters of \textit{Arabidopsis thaliana} on the Iberian Peninsula using a Logistic Normal Dirichlet Model for spatial compositional data (Section~\ref{sec:arabidopsis}), and modelling of wind speed data in the United Kingdom using spatio-temporal models for continuous domains with non-separable covariance structures (Section~\ref{sec:spacetime}). Finally, Section~\ref{sec:Discussion} ends with some conclusions.
The data and R code to reproduce the examples, simulation studies, and real data analysis presented in this paper are available at \url{https://github.com/spatialstatisticsupna/INLA_groupCV}.

\section{Leave-group-out cross-validation for latent Gaussian models} \label{sec:LGOCV}

LGOCV takes each data point, represented by $y_i$, as a testing set,
and creates an index set $I_i$ specific to each point based on the
prediction task. Using all observed data except the data indexed by
$I_i$, denoted by $\boldsymbol{y}_{-I_{i}}$, it computes the posterior predictive density
$\pi(Y_i|\boldsymbol{y}_{-I_{i}})$ and validates the prediction using
the observed data point $y_i$ and scoring rules. However, when the
model structure is complex or the prediction task is unclear,
constructing $I_i$ can be challenging. To overcome this, we can
automatically construct $I_i$ by selecting the most informative data
points to predict $y_i$. This automatic construction is motivated by
the fact that most extrapolation tasks have less information available
from the data used for prediction. While constructing $I_i$
automatically can be challenging in general model settings, it becomes
straightforward when the model is represented as a latent Gaussian
model. Furthermore, using a latent Gaussian model allows for fast and
accurate approximation of posterior densities when excluding partial
data.

A latent Gaussian model is composed of three layers that can be formulated by
\begin{equation*}
    \label{eq:lgm}
    \begin{split}
      &y_i|\eta_i,\boldsymbol{\theta} \sim \pi(y_i|\eta_i,\boldsymbol{\theta}), \\
      &\boldsymbol{\eta} = \boldsymbol{A}\boldsymbol{f},\\
      &\boldsymbol{f}|\boldsymbol{\theta} \sim N(0,\boldsymbol{P}_{\boldsymbol{f}}(\boldsymbol{\theta})),\\
      &\boldsymbol{\theta} \sim \pi(\boldsymbol{\theta}).
    \end{split}
\end{equation*}
The observed data $\boldsymbol{y}$ is modelled
conditionally independent given the hyperparameter
$\boldsymbol{\theta}$ and its corresponding linear predictor $\eta_i$
using the likelihood function $\pi(y_i|\eta_i,\boldsymbol{\theta})$.
The linear predictor $\boldsymbol{\eta}$ is a linear combination of
model effects $\boldsymbol{f}$, which has a multivariate Gaussian
prior with a zero mean and a precision matrix
$\boldsymbol{P}_{\boldsymbol{f}}(\boldsymbol{\theta})$ that depends on
$\boldsymbol{\theta}$. We assign a prior distribution to
$\boldsymbol{\theta}$. Using INLA method, we can have a Gaussian
approximation of the posterior density
$\pi(\boldsymbol{f}|\boldsymbol{\theta},\boldsymbol{y}_{-I_{i}})$.
The automatic groups are constructed using the correlation matrix of
$\boldsymbol{f}$, which can be derived from either the prior density
or the approximated posterior density at the mode of hyperparameter
configuration. See \cite{Liu2022} for further details.

We illustrate how to construct $I_i$ using a simple
example. Suppose we have correlation coefficients $\boldsymbol{C}_i$,
which contain the correlation of the linear predictor $\eta_i$ with
all $\eta_j$, for $j=1,\ldots,n$. If
$\boldsymbol{C}_i = \{1,1,0.9,0.9,0.8,0.8,-0.1,-0.1,0,0\}$, we define
$\boldsymbol{L}_i$ as the sorted unique absolute values of
$\boldsymbol{C}_i$, giving us
$\boldsymbol{L}_i = \{1.0,0.9,0.8,0.1,0.0\}$. We then include the most
correlated data points based on the $m$ largest correlation levels
in $I_i$. For example, if we choose $m = 3$, then
$I_i = \{1,2,3,4,5,6\}$. We refer to the parameter $m$ as the number of
level sets, which is determined by the degree of extrapolation implied
by LGOCV. When the information about the degree of extrapolation is
missing, an empirical suggestion is to use $m = 3$, or we could
compute a sequence of $m$ values to determine the predictive
performance.

The automatic group construction procedure for LGOCV is implemented in the R-INLA package through the \texttt{inla.group.cv()} function, whose input argument is a previously fitted \texttt{inla} model. The output of this function contains: (i) the automatically constructed groups $I_i$ by selecting the most informative data points to predict $y_i$ based on posterior correlations between the linear predictors (which can be used as an input parameter to compute predictive measures for other models), and (ii) the cross-validated predictive density $\pi(Y_i=y_i | \boldsymbol{y}_{-I_{i}})$ and additional statistics to compute $E[Y_i |\boldsymbol{y}_{-I_{i}}]$. The efficient implementation of the automatic group construction procedure allows computations on relatively large datasets in a few seconds.

\section{Evaluating cross-validation measures in structured models} \label{sec:Validation}

In the forthcoming examples, our focus is on analyzing the extrapolation performance of structured models, in which the unobserved data is often assumed less dependent on the observed data. For this end, we compare the behaviour of different models in terms of
(i) Bayesian model selection criteria such as the deviance information criterion (DIC; \citealp{Spiegelhalter2002}) and the Watanabe-Akaike information criterion (WAIC; \citealp{watanabe2010asymptotic}), and (ii) predictive performance measures based on LOOCV and the recently proposed automatic group construction procedure for LGOCV for latent Gaussian models using the well-known INLA estimation technique. Specifically, we compute commonly used utility functions such as the
logarithmic score \citep{gneiting2007strictly}
\begin{equation*}
    \mbox{LS} = - \frac{1}{n}\sum\limits_{i=1}^n \log \pi(Y_i=y_i | \boldsymbol{y}_{-I_{i}}),
\end{equation*}
and the mean square prediction error
\begin{equation*}
    \mbox{MSPE} = \frac{1}{n}\sum\limits_{i=1}^n \left(E[Y_i |\boldsymbol{y}_{-I_{i}}]-y_i \right)^2,
\end{equation*}
where $\pi(Y_i=y_i | \boldsymbol{y}_{-I_{i}})$ denotes the cross-validated predictive density at the observed data $y_i$.
Note that in the case of LOOCV the training set $\boldsymbol{y}_{-I_{i}}$ reduces to ${\y}_{-i}=(y_1,\ldots,y_{i-1},y_{i+1},y_n)^{'}$. See \cite{Liu2022} for details about the computation of $E[Y_i |\boldsymbol{y}_{-I_{i}}]$.

In our first example, we compare the predictive performance exhibited by different temporal models for long-term forecasting in time series data. Our second example delves into the extrapolation performance of spatial disease mapping models for areal count data, that is, predictions in non-observed spatial units. We show that DIC, WAIC and leave-one-out cross-validation (LOOCV) measures points to different models from those selected by using utility (loss) functions based on the recently proposed automatic LGOCV. We also validate the predictive performance of these models by imitating real extrapolation tasks for both examples based on temporal and spatial data.
All computations are made using the recently implemented ``compact'' mode \citep{van2023new} in \texttt{R-INLA} (stable) version INLA\_23.09.09 of R-4.3.1.

\subsection{Example 1: temporal models} \label{sec:Example_TemporalData}

In this example, we want to reproduce a real situation where the aim is to perform long-term forecasting through the use of temporal models. Let us consider a scenario where our interest lies in forecasting surface temperatures using shortwave radiation as a predictor variable.
We start by simulating radiation data $X$ for a total of $n=2000$ time points using a first-order autoregressive (AR1) model, where the autocorrelation coefficient is set at $\rho=0.9$ and the marginal variance equals to one. Then, we generate values of surface temperature $Y$ using the following linear model:
\begin{equation*}
    \label{eq:Example1}
    Y = \beta_0 + X + e, \quad \mbox{with} \quad e \sim N(0,0.1).
\end{equation*}
Finally, let us assume that $\tilde{X}=X \cdot |X|$ corresponds to an imperfect observation of the shortwave radiation $X$ (arising from potential instrumental deficiencies, for instance).
The definition of $\tilde{X}$ introduces a method for establishing a non-linear relationship between the response variable (temperature) and the observed covariate, resembling the true underlying linear association with the predictor variable.

Two different models are considered to estimate the surface temperature $y_t$ based on observed radiation data $\tilde{x}_t$ for $t=1,\ldots,2000$ time points.
\begin{itemize}[leftmargin=0pt]
\item[] \textbf{\underline{Linear model}:}
    \begin{equation*}
        \begin{split}
        &y_t | \eta_t,\boldsymbol{\theta} \sim N(\eta_t,\tau_y), \\
        &\eta_t = \beta_0 + \beta \cdot \tilde{x}_t,
        \end{split}
    \end{equation*}
where $\beta_0$ is an intercept and $\beta$ is the fixed effect coefficient associated to the covariate $\tilde{x}_t$.

\item[] \textbf{\underline{AR1 model}:}
    \begin{equation*}
        \begin{split}
        &y_t | \eta_t,\boldsymbol{\theta} \sim N(\eta_t,\tau_y), \\
        &\eta_t = \beta_0 + u_t,\\
        &u_t = \rho \cdot u_{t-1} + \epsilon_t,\\
        &\epsilon_t \sim N(0,\tau_u)
        \end{split}
    \end{equation*}
where the temporally structured random effect ${\u}=(u_1,\ldots,u_{2000})^{'}$ with AR1 prior distribution is used to model the surface temperature as a substitute of the observed predictor $\tilde{x}_t$.

    %
\end{itemize}

\medskip

\begin{figure}[!ht]
    \begin{center}
        \vspace{-0.2cm}
        \includegraphics[width=0.7\textwidth]{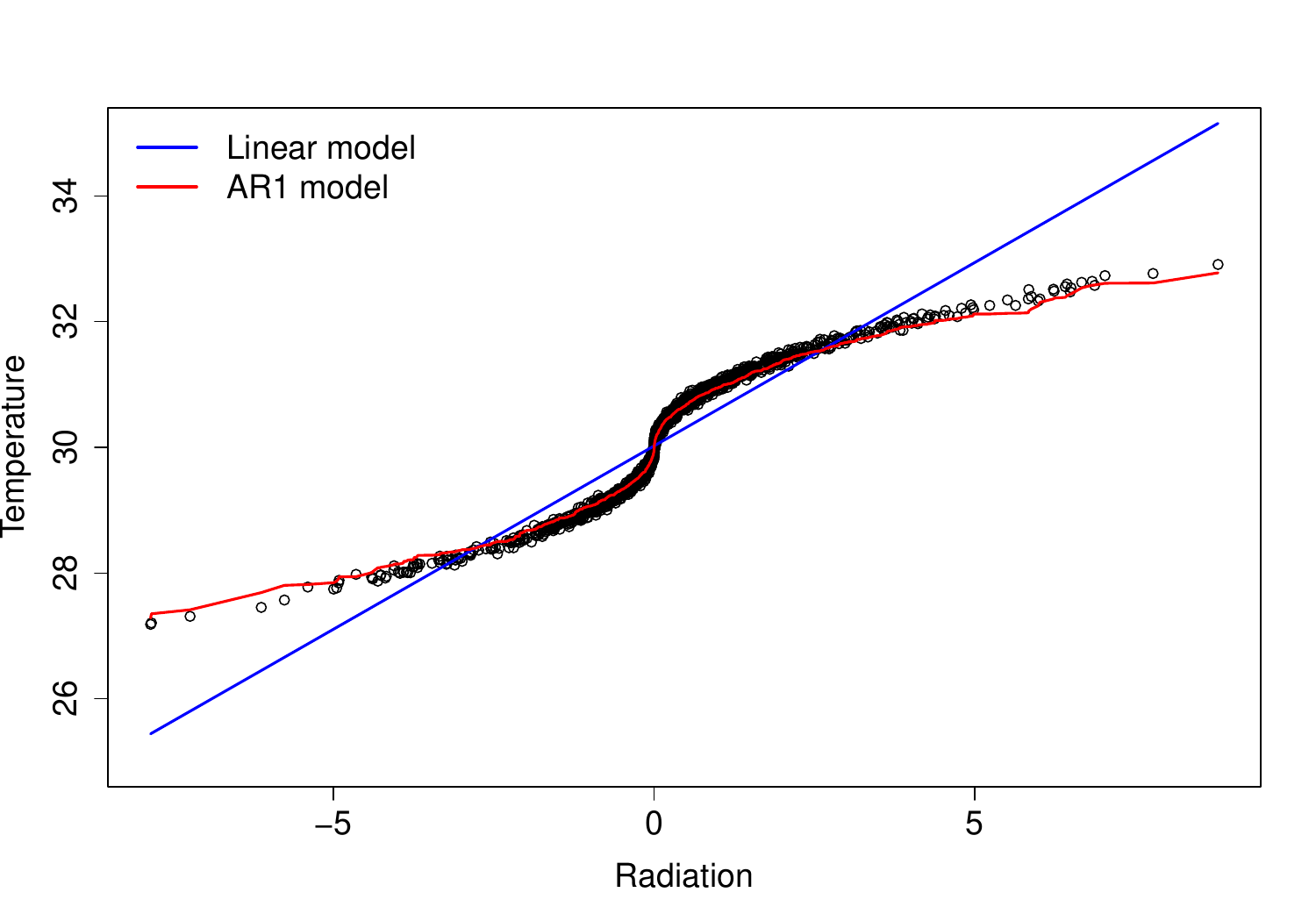}
        \vspace{-0.5cm}
    \end{center}
    \caption{Dependence structure captured between the observed
        shortwave radiation $\tilde{X}$ and estimated surface
        temperature $\hat{Y}$.}
    \label{fig:TemporalData_tempVSradiation}
\end{figure}

\begin{figure}[!ht]
    \begin{center}
        \includegraphics[width=\textwidth]{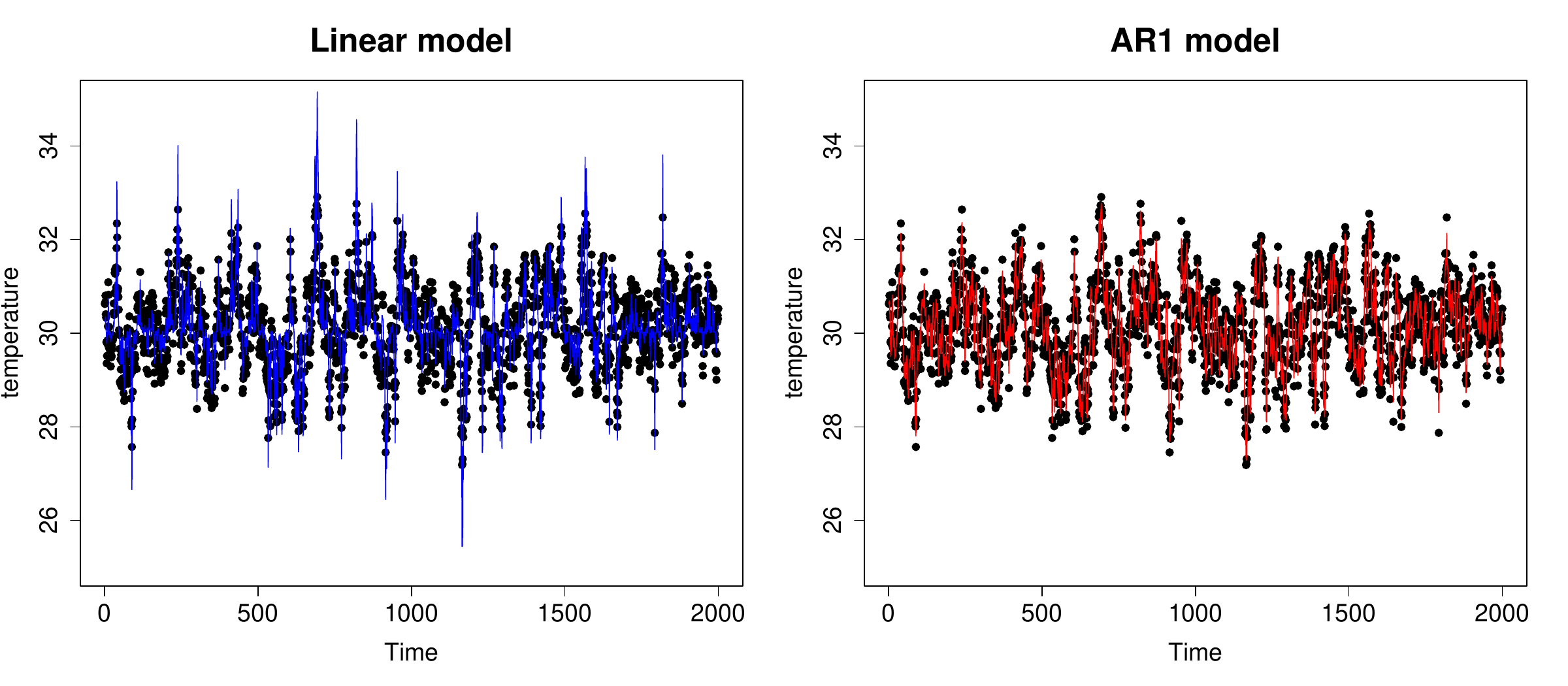}
    \vspace{-1cm}
    \end{center}
    \caption{True values (black dots) and posterior median estimates
        (colored lines) for the surface temperature
        ${\y}_t=(y_1,\ldots,y_{2000})^{'}$.}
    \label{fig:TemporalData_ModelFitting}
\end{figure}

\autoref{fig:TemporalData_tempVSradiation} shows how the models capture the underlying dependence structure between the observed radiation and temperature values. Clearly, the AR1 model better captures the non-linear relationship between these variables. Posterior median estimates for the temperature values ${\y}_t = (y_1, \ldots, y_{2000})^{'}$ obtained with both models are plotted in \autoref{fig:TemporalData_ModelFitting}.

\autoref{tab:Example_TemporalData} displays the values of model selection criteria and predictive performance measures (logarithmic score and mean square prediction error) for LOOCV and LGOCV with automatic groups construction (m=3), derived from either the prior precision matrix or the posterior correlation matrix of each model. See \cite{Liu2022} for further details.
The groups derived from the AR1 model (as this is the true generating model) are used as a reference to make predictive measures comparable. As expected, the AR1 model is pointed out as the best model in terms of model fitting and predictive performance based on the LOOCV approach. However, predictive measures based on LGOCV points to the linear model. 
It should be noted that in temporally structured models, the LOOCV technique essentially evaluates their interpolation performance, i.e., how effectively the model fills in the missing values within the observed time series.
In contrast, LGOCV assesses the model's capability when the prediction task is to forecast the response variable for future time points.

\begin{table}[!ht]
\caption{Model comparison for temporal data: model selection
        criteria (posterior mean deviance $\bar{D}$, effective number
        of parameters $p_D$, DIC and WAIC) and predictive
        performance measures for LOOCV and LGOCV with automatic groups
        construction (m=3) based on prior and posterior correlations
        for each temporal model.}
    \label{tab:Example_TemporalData}
    \begin{center}
        \resizebox{\textwidth}{!}{
            \begin{tabular}{l|rrrr|rr|rr|rr}
              \toprule
              & \multicolumn{4}{c|}{\multirow{2}{*}{\bf Model selection criteria}} & \multicolumn{2}{c|}{\multirow{2}{*}{\bf LOOCV}} & \multicolumn{4}{c}{\bf LGOCV}\\
              & & & & & & & \multicolumn{2}{c|}{\texttt{"prior"}} & \multicolumn{2}{c}{\texttt{"posterior"}} \\
              Model & $\bar{D}$ & $p_D$ & DIC & WAIC & LS & MSPE & LS & MSPE & LS & MSPE \\[0.5ex]
              \hline
              Linear      &  1500.0 &    3.0 &  1503.0 &  1507.7 &  0.376 & 0.125 &  0.383 & 0.126 &  0.383 & 0.126 \\
              AR1         &  -439.5 & 1286.7 &   847.2 &   622.3 &  0.349 & 0.116 &  0.824 & 0.303 &  0.823 & 0.302 \\
              \bottomrule
            \end{tabular}}
    \end{center}
\end{table}

To validate these measures, we conduct the following simulation study. We start by considering our training set as the time series
encompassing the period $t=\{1,\ldots,1900\}$ and compute long-term predictions for the next 100 time points using INLA. We repeat this process by shifting the time window one step backward for a total of 500 data sets. That is, our last historic data corresponds to the period $t=\{1,\ldots,1400\}$, so that predictions are computed for future time points $\{1401,\ldots,1500\}$. We compute the mean absolute prediction errors (MAPE) and root mean square prediction errors (RMSPE) between the posterior median estimates and true values of temperature for each data set, $s=1,\ldots,500$, as
\begin{equation*}
    \mbox{MAPE}_s = \frac{1}{100} \sum\limits_{j=1}^{100} \left|\hat{y}_{t+j} - y_{t+j} \right|
    \qquad \mbox{and} \qquad
    \mbox{RMSPE}_s = \sqrt{\frac{1}{100} \sum\limits_{j=1}^{100} (\hat{y}_{t+j} - y_{t+j})^2}.
\end{equation*}

\begin{figure}[!ht]
    \begin{center}
        \includegraphics[width=\textwidth]{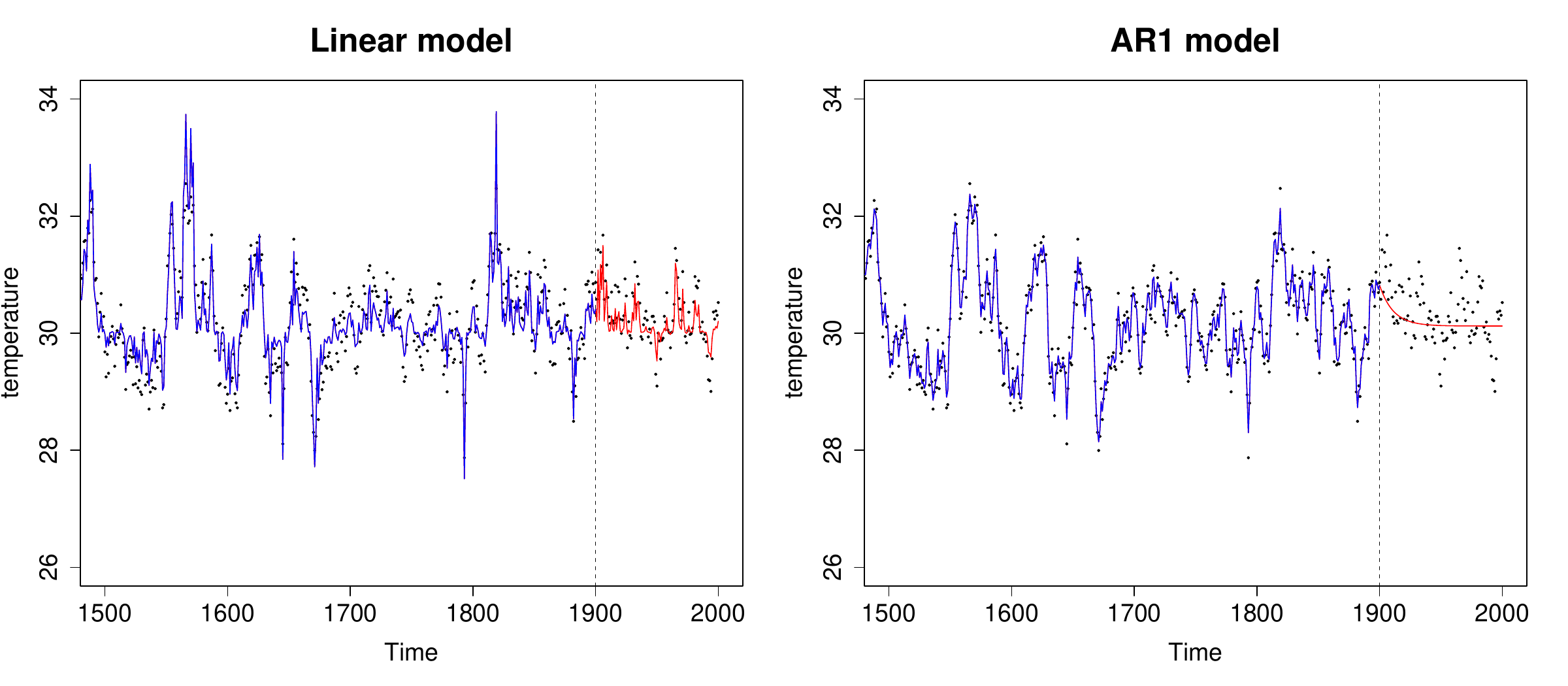}
        \vspace{-1cm}
    \end{center}
    \caption{True values (black dots), posterior median estimates
        (blue line) and forecasted values (red lines) obtained from
        linear and AR1 models for the first data set of the simulation
        study.}
    \label{fig:TemporalData_Predictions}
\end{figure}

\autoref{tab:TemporalData_Predictions} presents the average values of these metrics calculated across the 500 data sets. Our simulation study corroborates that the linear model outperforms the AR1 model when the objective is to perform long-term forecasting. This is something expected, as the linear model incorporates additional auxiliary information derived from the observed covariate, resulting in more accurate predictions. Conversely, the AR1 model converges toward the overall mean as time advances, as illustrated in \autoref{fig:TemporalData_Predictions}.

\begin{table}[!bt]
    \caption{Average values over 500 data sets of MAPE and RMSPE for
        long-term predictions.}
    \label{tab:TemporalData_Predictions}
    \begin{center}
        \begin{tabular}{l|rr}
          \toprule
          Model & MAPE & RMSPE \\[0.5ex]
          \hline
          Linear  & 0.297 & 0.334 \\
          AR1     & 0.653 & 0.800 \\
          \bottomrule
        \end{tabular}
    \end{center}
\end{table}

\subsection{Example 2: spatial models} \label{sec:Example_SpatialData}

The objective of this example is to validate the effectiveness of automatic groups construction within LGOCV for extrapolation prediction tasks in disease mapping models, that is, when the goal is to predict disease risks or rates in unobserved regions using spatially structured models.

\begin{figure}[!ht]
    \begin{center}
        \includegraphics[width=1.3\textwidth]{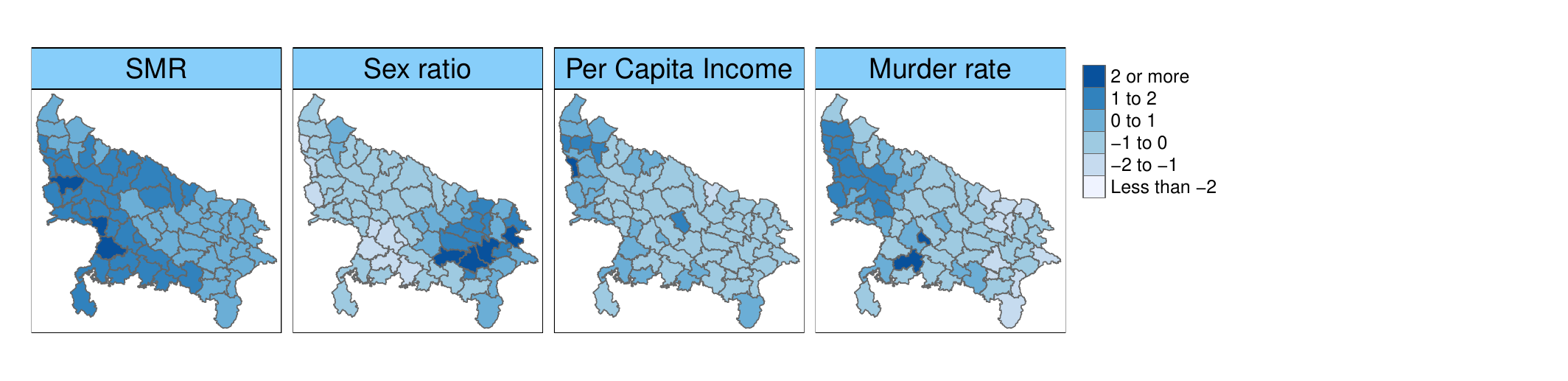}
        \vspace{-1.5cm}
    \end{center}
    \caption{Maps of the standardized mortality ratio (SMR) of dowry
        deaths in the districts of Uttar Prades and standardized
        potential risk factors.}
    \label{fig:SpatialData_MapCovariates}
\end{figure}

We use data concerning dowry deaths across the 70 districts of Uttar Pradesh (the most populous state in India) during the year of 2011, a form of crime against women which is very specific to India. For each district, we have records of the number of observed and expected dowry deaths.  The expected deaths are calculated by considering women aged 15 to 49 years as the population at risk. In addition, we also consider the following predictors as potential areal risk factors: sex ratio (number of females per 1,000 males), per capita income, and number of murders per 100,000 inhabitants. In \autoref{fig:SpatialData_MapCovariates} we plot the maps of the standardized mortality ratio (SMR) and standardized covariates at small area level.
Similar data were analyzed by \cite{vicente2020crime} and \cite{adin2023alleviatting} to study the association between these
potential risk factors and dowry deaths in Uttar Pradesh during the period 2001-2011.

Let $y_i$ and $E_i$ be the number of observed and expected deaths, respectively. Conditional on the relative risks $r_i$, the
number of observed cases is assumed to follow a Poisson distribution
\begin{equation*}
    \begin{split}
    &y_i | r_i,\boldsymbol{\theta} \sim \mbox{Poisson}(\mu_i=E_i r_i) \\
    &\log{\mu_i} = \log{E_i} + \log{r_i}
    \end{split}
\end{equation*}
where $\log{E_i}$ is an offset and depending on the specification of $\eta_i=\log{r_i}$ different models are defined. Here, different spatially structured models are considered to estimate the relative risks of dowry deaths for each small area $i=1,\ldots,70$.
\begin{itemize}[leftmargin=0pt]
\item[] \textbf{\underline{CAR model}:} $$\eta_i = \beta_0 + \xi_i,$$
where $\beta_0$ is an intercept representing the overall log-risk and $\xi_i$ is a spatially structured random effect for which the so-called BYM2 conditional autoregressive (CAR) prior distribution \citep{riebler2016intuitive} is assumed.
\item[] \textbf{\underline{CAR model + covariates}:} $$\eta_i = \beta_0 + x_i^{'}{\bbeta} + \xi_i,$$
where $x_i^{'}$ is the vector of standardized covariates in the $i$-th small area and ${\bbeta}=(\beta_1,\beta_2,\beta_3)^{'}$ is the vector of fixed effects coefficients.
\item[] \textbf{\underline{P-spline model}:} $$\eta_i = \beta_0 + f(s_{1i},s_{2i}),$$
where $f(s_{1i},s_{2i})$ is a spatial smooth function that is approximated as $f({\s}_1,{\s}_2) \approx {\B}_s {\ttheta}_s$,
where ${\B}_s = {\B}_2 \Box {\B}_1$ is the two-dimensional B-spline basis of dimension $70 \times k_s$ (with $k_s=k_1k_2$
depending on the number of knots and the degree of the polynomials in the bases ${\B}_1$ and ${\B}_2$), obtained from the row-wise
Kronecker product of the marginal bases for longitude ${\s}_1 = (s_{1,1},\ldots,s_{1,70})^{'}$ and latitude ${\s}_2 = (s_{2,1},\ldots,s_{2,70})^{'}$. To achieve smoothness, the following prior distribution is assumed for the unknown coefficients ${\ttheta}_s=(\theta_1,\ldots,\theta_{k_s})^{'}$
\begin{equation*}
    {\ttheta}_s \sim N({\0},{\PP}_s^{-})
\end{equation*}
where ${\PP}_s$ is a spatial anisotropic precision matrix that induces different amount of smoothing in longitude an latitude
dimensions (see \citealp{ugarte2017one}). The symbol $^{-}$ denotes the Moore-Penrose generalized inverse of a matrix.
\item[] \textbf{\underline{P-spline model + covariates}:}
    $$\eta_i= \beta_0 + x_i^{'}{\bbeta} + f(s_{1i},s_{2i})$$
\end{itemize}

\begin{figure}[!ht]
    \begin{center}
        \includegraphics[width=1.25\textwidth]{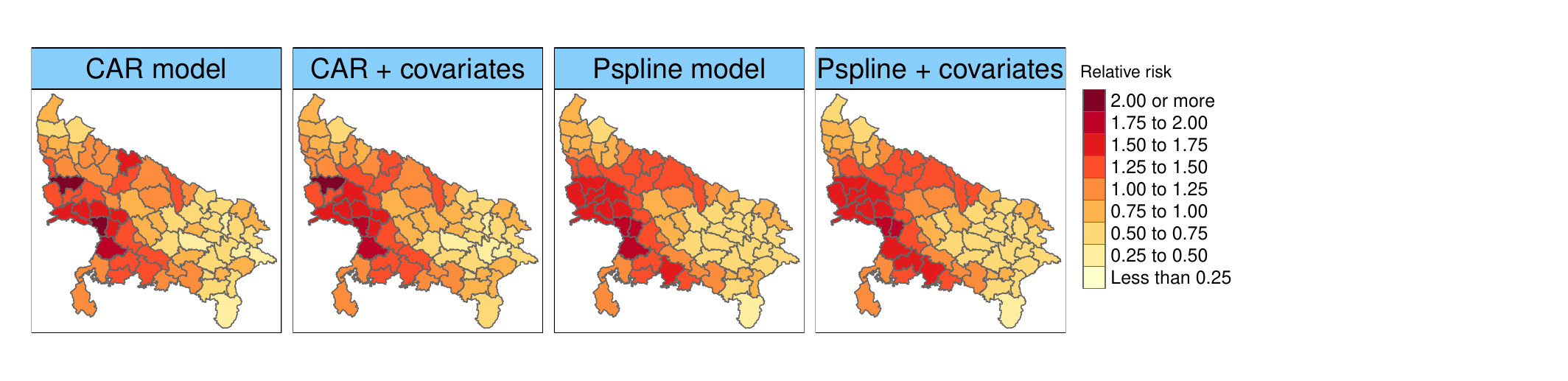}
        \vspace{-1.5cm}
    \end{center}
    \caption{Maps of posterior median estimates of dowry death
        relative risks obtained from the different spatial models.}
    \label{fig:SpatialData_MapRisks}
\end{figure}

The maps illustrating the posterior median estimates of $r_i=\exp(\eta_i)$ are presented in \autoref{fig:SpatialData_MapRisks}. In general, similar geographical distributions of mortality risks related to dowry deaths are observed across the different spatial models.
\autoref{tab:Example_SpatialData} compares the values of model selection criteria and predictive performance measures considering LOOCV and LGOCV with automatic groups construction (m=3) based on posterior correlations for each spatial model. According to DIC/WAIC measures, CAR models are preferred over P-spline models in terms of model selection criteria. In addition, predictive measures derived from LOOCV suggest that the incorporation of spatial covariates into the linear predictor of the CAR/P-spline models does not enhance their predictive performance. Considering these measure, the P-spline model with covariates exhibits the poorest performance in both the trade-off between model fit and complexity, and in terms of predictive accuracy.
In contrast, the predictive metrics obtained through LGOCV indicate that the CAR models perform less effectively in predicting relative risks for unobserved small areas. This is something expected considering that CAR models induce local correlations among neighbouring areas, while P-spline models are able to borrow spatial information for areas that are further apart. Similar conclusions are obtained for higher values of $m$.

\begin{table}[!ht]
    \caption{Model comparison for spatial data: model selection
        criteria (posterior mean deviance $\bar{D}$, effective number
        of parameters $p_D$, DIC and the WAIC) and predictive
        performance measures for LOOCV and LGOCV with automatic groups
        construction (m=3) based on posterior correlations for each
        spatial model.}
    \label{tab:Example_SpatialData}
    \begin{center}
        \resizebox{\textwidth}{!}{
        \begin{tabular}{l|rrrr|rr|rr}
          \toprule
          & \multicolumn{4}{c|}{\bf Model selection criteria} & \multicolumn{2}{c|}{\bf LOOCV} & \multicolumn{2}{c}{\bf LGOCV}\\[0.5ex]
          Model & $\bar{D}$ & $p_D$ & DIC & WAIC & LS & MSPE & LS & MSPE \\[0.5ex]
          \hline
          CAR                & 433.2 & 47.5 & {\bf 480.7} & {\bf 473.4} & {\bf 3.604} &  {\bf 90.529} & 3.611 & 89.480 \\
          CAR+covariates     & 435.4 & 45.6 & 481.1 & 476.8 & 3.616 & 105.042 & 3.614 & 99.982 \\
          Pspline            & 454.0 & 30.4 & 484.4 & 492.0 & 3.600 &  93.196 & {\bf 3.485} & {\bf 76.146} \\
          Pspline+covariates & 459.5 & 30.8 & 490.3 & 501.0 & 3.667 & 101.102 & 3.508 & 82.271 \\
          \bottomrule
        \end{tabular}}
    \end{center}
\end{table}

\begin{figure}[!ht]
    \begin{center}
        \vspace{0.5cm}
        \includegraphics[width=0.9\textwidth]{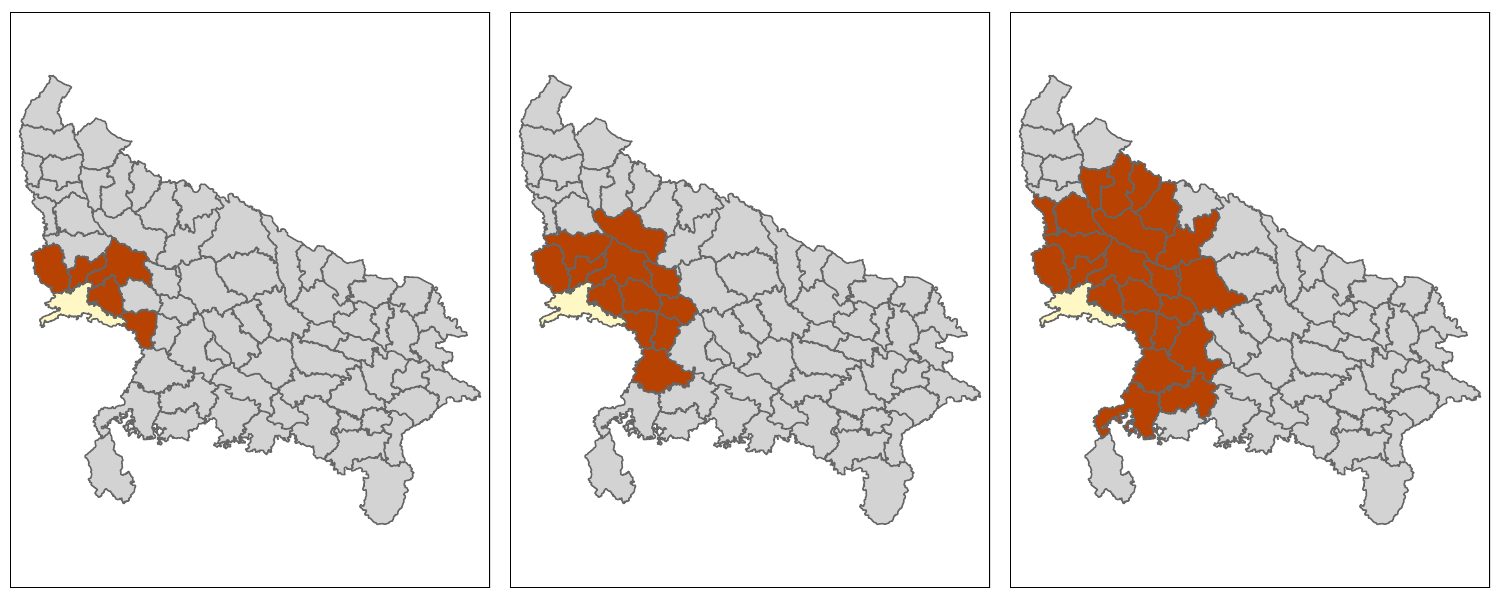}
        \vspace{-0.5cm}
    \end{center}
    \caption{Maps with 1st-order (left), 2nd-order (center) and
        3rd-order (right) neighbourhood areas for a selected district
        of Uttar Pradesh.}
    \label{fig:SpatialData_SimulationStudy}
\end{figure}

As in the previous example, we should notice that LOOCV is not an appropriate approach to estimate the models predictive performance
when the task is to predict relative risks in unobserved group of areas (extrapolation task). This concern is particularly relevant when employing spatially structured models, where the linear predictor of each area is highly correlated with those of its neighbouring regions.
Once again, our focus lies in validating these measures through a simulation study. For each small area, we remove its k-order neighbouring or spatially contiguous areas ($k=1,2$ and 3) and compute posterior predictive distributions for the relative risks on those areas under the different models fitted with INLA. In \autoref{fig:SpatialData_SimulationStudy} we show a map with the 1st, 2nd and 3rd-order neighbourhood areas for a randomly selected district of Uttar Pradesh.
In \autoref{tab:SpatialData_Predictions} we show average values of mean relative prediction errors (MRPE) and relative root mean square
prediction errors (RRMSPE) computed over each data set. As expected, higher MRPE and RRMSPE values are obtained as the number of removed areas increases, which is controlled by the parameter $k$. Our simulation study corroborates that, as pointed out by the LGOCV approach, CAR models exhibit inferior predictive performance compared to P-spline models in the context of spatial extrapolation. In general, we observe that better predictive count estimates are obtained when including auxiliary information of area-level covariates as predictors in the models.

\begin{table}[!ht]
    \caption{Average values over 70 data sets of MRPE and RRMSPE for
        predicted counts.}
    \label{tab:SpatialData_Predictions}
    \begin{center}
        \begin{tabular}{l|ccc|ccc}
          \toprule
          & \multicolumn{3}{c|}{MRPE} & \multicolumn{3}{c}{RRMSPE} \\
          & & & & & & \\[-2.ex]
          Model & k=1 & k=2 & k=3 & k=1 & k=2 & k=3 \\[0.5ex]
          \hline
          CAR                  & 0.273 & 0.325 & 0.370 & 0.354 & 0.442 & 0.521 \\
          CAR + covariates     & 0.281 & 0.299 & 0.310 & 0.357 & 0.395 & 0.410 \\
          Pspline              & 0.270 & 0.316 & 0.344 & 0.347 & 0.422 & 0.470 \\
          Pspline + covariates & 0.161 & 0.171 & 0.172 & 0.215 & 0.246 & 0.253 \\
          \bottomrule
        \end{tabular}
    \end{center}
\end{table}

\pagebreak
\section{Applications} \label{sec:Applications}

When constructing groups for different models, it is essential to choose a reference model to ensure that we can obtain comparable measures using the LGOCV approach. The simulation study presented as Supplementary Material suggests that automatic groups generated from the posterior of a suitable model fitted to a dataset is the recommended approach for evaluating various candidate models. Based on these results, for the real data analyses presented in this section, we decided to use models with the lowest DIC values as the reference models for constructing the automatic groups and subsequently comparing the predictive performance of the different candidates.

The supplementary material also contains additional figures pertaining to the real data applications on spatial statistics discussed in the subsequent sections.

\subsection{Joint modelling of pancreatic cancer mortality and incidence data}
\label{sec:DiseaseMapping}

Bayesian hierarchical models are a powerful tool for analyzing area-level incidence or mortality data, providing valuable insight into latent spatial, temporal, and spatio-temporal patterns. These models typically rely on generalized linear mixed models that incorporate spatially and temporally structured random effects, allowing for the smoothing of disease risks or rates by incorporating information from neighboring areas and time periods. 
%
%
Further extensions of spatio-temporal disease mapping models aim to jointly analyze several responses by the construction of multivariate proposals based on Gaussian Markov random fields. An exhaustive review on the topic can be found in the work by \cite{MacNab2018Test}. Alternatively, shared-component models have been widely used to investigate closely related phenomena or diseases that share a recognized common risk factor (see \citealp{knorr2001shared,held2005towards}).

\subsubsection{The models}
In this section, we demonstrate the use of the automatic group construction procedure for LGOCV when the objective is to jointly model incidence and mortality data for a highly lethal disease such as pancreatic cancer using different spatio-temporal models. Specifically, we conduct an analysis using yearly count data for the male population across the 105 Clinical Commissioning Groups (CCG) of England during the period 2001-2020 (official data provided by the National Health Service in England available at \url{https://www.cancerdata.nhs.uk/incidence_and_mortality}). In these types of studies, we expect that using multivariate models will enhance the performance of individual models by leveraging the strong correlation between incidence and mortality data.

The setting for our study is the following. Let us define as $y_{it1}$ and $y_{it2}$ the number of mortality and incidence cases, respectively, by health-area (CCG) $i=1,\ldots,S=105$ and time period $t=1,\ldots,T=20$ (corresponding to the period 2001-2020). Conditional on the relative risks $r_{itj}$ for mortality ($j=1$) and incidence ($j=2$) data, we assume that the number of observed cases follow a Poisson distribution with mean $\mu_{itj}=E_{itj} \cdot r_{itj}$, that is,
\begin{equation*}
    \begin{split}
    & y_{itj} | r_{itj}, \boldsymbol{\theta} \sim \mbox{Poisson}(\mu_{itj}=E_{itj} \cdot r_{itj}), \\
    & \log{\mu_{itj}} = \log{E_{itj}} + \log{r_{itj}}.
    \end{split}
\end{equation*}
Here $E_{itj}$ is computed using indirect standardization as $E_{itj}=\sum_k n_{itjk} \cdot m_{jk}$, where $k$ denotes the age-groups ($<25$, $[25,50)$, $[50,60)$, $[60,70)$, $[70,80)$ and $\geq 80$), $n_{itjk}$ is the population at risk in the area $i$, time $t$ and age-group $k$, and $m_{jk}$ is the overall mortality/incidence rate in England during the whole study period for the $k$-th age-group. Then, the log-risk is modelled as
\begin{equation}
\label{eq:DiseaseMapping}
\log r_{itj} = \alpha_j + \phi_{ij} + \gamma_{tj} + \delta_{itj},
\end{equation}
where $\alpha_j$ is an intercept representing the overall log-risk, $\phi_{ij}$ and $\gamma_{tj}$ are the spatial and temporal main effects, and $\delta_{itj}$ is the spatio-temporal interaction for mortality and incidence data respectively.

Three different disease mapping models are considered to jointly estimate pancreatic cancer mortality and incidence data under the model formulation of Equation~\eqref{eq:DiseaseMapping}. Firstly, independent spatio-temporal models (M1) are considered. For comparison purposes with multivariate models, if we denote as ${\r}=({\r}_1^{'},{\r}_2^{'})^{'}$ to the joined vector of mortality and incidence risks with ${\r}_j=(r_{11j},\ldots,r_{S1j},\ldots,r_{1Tj},\ldots,r_{STj})^{'}$, we formulate the model in matrix form as
\begin{equation}
\label{eq:DM_M1}
\mbox{\footnotesize\(
\begin{pmatrix} \log {\r}_1 \\ \log {\r}_2 \end{pmatrix} =
({\I}_2 \otimes {\1}_{T} \otimes {\1}_{S}) \begin{pmatrix} \alpha_1 \\ \alpha_2 \end{pmatrix} +
({\I}_2 \otimes {\1}_{T} \otimes {\I}_S) \begin{pmatrix} {\pphi}_1 \\ {\pphi}_2 \end{pmatrix} +
({\I}_2 \otimes {\I}_{T} \otimes {\1}_S) \begin{pmatrix} {\ggamma}_1 \\ {\ggamma}_2 \end{pmatrix} +
({\I}_2 \otimes {\I}_{T} \otimes {\I}_{S}) \begin{pmatrix} {\ddelta}_1 \\ {\ddelta}_2 \end{pmatrix},
\)}
\end{equation}
where ${\1}_S$ and ${\1}_T$ are column vectors of ones of length $S$ and $T$ respectively, ${\I}_2$, ${\I}_S$ and ${\I}_T$ are identity matrices of dimension $2 \times 2$, $S \times S$ and $T \times T$ respectively, ${\pphi}_j=(\phi_{1j},\ldots,\phi_{Sj})^{'}$ is a spatial random effect with intrinsic CAR prior distribution, ${\ggamma}_j=(\gamma_{1j},\ldots,\gamma_{Tj})^{'}$ is a temporally structured random effect that follows a first order RW prior distribution, and ${\ddelta}_j=(\delta_{11j},\ldots,\delta_{STj})^{'}$ is a spatio-temporal random effect where the four different types of interactions originally proposed by \cite{knorr2000} are considered.
That is, for $j=1,2$
%
%
\begin{equation*}
{\pphi}_j \sim N({\0},[\tau_{\phi_j}{\RR}_{\phi}]^{-}),
\qquad
{\ggamma}_j \sim N({\0},[\tau_{\gamma_j}{\RR}_{\gamma}]^{-}),
\qquad \mbox{and} \qquad
{\ddelta}_j \sim N({\0},[\tau_{\delta_j}{\RR}_{\delta}]^{-}),
\end{equation*}
where ${\RR}_{\phi}$, ${\RR}_{\gamma}$ and ${\RR}_{\delta}$ are structure matrices for the corresponding spatial, temporal and spatio-temporal random effects respectively, and $\tau_{\phi_j}$, $\tau_{\gamma_j}$ and $\tau_{\delta_j}$ are precision parameters. See, for example, \cite{goicoa2018spatio} for details about identifiability issues in this type of spatio-temporal models and its implementation using INLA.

Alternatively, two distinct multivariate models that enable the incorporation of correlation structures between mortality and incidence data are considered: a spatio-temporal extension of the so-called M-based model described in \cite{vicente2020bayesian} (M2) and models including shared-component terms for space and time similar to the ones described in \cite{etxeberria2023using} (M3).
Denote as ${\pphi}=({\pphi}_1^{'},{\pphi}_2^{'})^{'}$ and ${\ggamma}=({\ggamma}_1^{'},{\ggamma}_2^{'})^{'}$ the joined vectors of spatial and temporal random effects for mortality and incidence risks, respectively. Then, the M-model formulation (M2) is identical to the expression in Equation~\eqref{eq:DM_M1}, assuming the following prior distribution
\begin{equation*}
{\pphi} \sim N\left({\0}, {\SSigma}_{\phi}^{-1} \otimes {\RR}_{\phi}^{-} \right),
\qquad \mbox{and} \qquad
{\ggamma} \sim N\left({\0}, {\SSigma}_{\gamma}^{-1} \otimes {\RR}_{\gamma}^{-} \right),
\end{equation*}
where ${\SSigma}_{\phi}$ and ${\SSigma}_{\gamma}$ are the covariance matrices between the spatial and temporal effects, respectively, of mortality and incidence data. See \cite{adin2023multivariate} for details about model implementation in INLA.
Finally, we formulate our shared component model (M3) as
\begin{equation*}
    \begin{split}
    & \log r_{it1} = \alpha_1 + c_{\phi}\phi_i + c_{\gamma}\gamma_t + \delta_{it1}, \\[1.ex]
    & \log r_{it2} = \alpha_2 + \dfrac{1}{c_{\phi}}\phi_i + \dfrac{1}{c_{\gamma}}\gamma_t + \delta_{it2},
    \end{split}
\end{equation*}
where $\phi_i$ is a shared spatial effect with intrinsic CAR prior distribution, $\gamma_t$ is a shared temporal effect with first order RW prior distribution, and $\delta_{itj}$ is a specific spatio-temporal random effect to model the interaction in mortality and incidence log-risks, respectively. Finally, $c_{\phi}$ and $c_{\gamma}$ are scaling parameters to allow a different ``risk gradient'' (on the log-scale) to be associated with the spatial and temporal components, respectively, for both responses. A summary with the main prior distributions for the three models considered in this section has been included in \autoref{tab:DM_models}.

\begin{table}[!t]
\begin{center}
\renewcommand*{\arraystretch}{1}
\caption{\label{tab:DM_models} Summary of the prior distributions for the spatial, temporal and spatio-temporal random effects under the three models considered in Section~\ref{sec:DiseaseMapping}.}
\vspace{0.2cm}
\resizebox{\textwidth}{!}{
\begin{tabular}{c|c|c|c}
\hline\\[-2ex]
& {\bf M1} & {\bf M2} & {\bf M3} \\
& (Independent models) & (M-model) & (Shared-component model) \\
\hline & & & \\[-1.5ex]
Spatial effect
& ${\pphi}_j \sim N({\0},[\tau_{\phi_j}{\RR}_{\phi}]^{-}), \quad j=1,2 $
& $\begin{pmatrix} {\pphi}_1 \\ {\pphi}_2 \end{pmatrix} \sim N\left({\0}, {\SSigma}_{\phi}^{-1} \otimes {\RR}_{\phi}^{-}\right)$
& ${\pphi} \sim N({\0},[\tau_{\phi}{\RR}_{\phi}]^{-})$ \\[2.5ex]
\hline & & & \\[-1.5ex]
Temporal effect
& ${\ggamma}_j \sim N({\0},[\tau_{\gamma_j}{\RR}_{\gamma}]^{-}), \quad j=1,2$
& $\begin{pmatrix} {\ggamma}_1 \\ {\ggamma}_2 \end{pmatrix} \sim N\left({\0}, {\SSigma}_{\gamma}^{-1} \otimes {\RR}_{\gamma}^{-}\right)$
& ${\ggamma} \sim N({\0},[\tau_{\gamma}{\RR}_{\gamma}]^{-})$ \\[2.5ex]
\hline \\[-1ex]
\multirow{2}{*}{Spatio-temporal}
& \multicolumn{3}{c}{\large ${\ddelta}_j \sim N({\0},[\tau_{\delta_j}{\RR}_{\delta}]^{-}) \quad j=1,2$}
\\[1.5ex]
\multirow{2}{*}{effect}
& \multicolumn{3}{c}{$\mbox{Type I: }{\RR}_{\delta}={\I}_T \otimes {\I}_S, \qquad \mbox{Type II: }{\RR}_{\delta}={\RR}_{\gamma} \otimes {\I}_S$}
\\[1.5ex]
& \multicolumn{3}{c}{$\mbox{Type III: }{\RR}_{\delta}={\I}_T \otimes {\RR}_{\phi}, \qquad \mbox{Type IV: }{\RR}_{\delta}={\RR}_{\gamma} \otimes {\RR}_{\phi}$}
\\[1.5ex]
\hline
\end{tabular}}
\end{center}
\end{table}

To fit the models, improper uniform prior distributions are given to all the standard deviations (square root inverse of precision parameters), and a Gamma(10,10) distribution is considered for the scaling parameters $c_{\phi}$ and $c_{\gamma}$ of the shared-component model. Finally, a vague zero mean normal distribution with a precision close to zero (0.001) is given to the model intercepts. 

\subsubsection{Results}
\autoref{tab:CancerData_DIC} presents a comparison of the different models used to estimate pancreatic cancer mortality and incidence relative risks in terms of model selection criteria and predictive performance measures. Both DIC/WAIC and predictive measures using a LOOCV approach consistently indicate that Type II interaction models should be selected. Among the considered models, the shared-component model M3 exhibits a slightly better level of performance. To compute comparable LS measures from the posterior predictive density functions $\pi(Y_{itj}|\boldsymbol{y}_{-I_{itj}})$, we use the groups $I_{itj}$ derived from Type II interaction models with number of level sets $m=5$. As shown in \autoref{tab:CancerData_DIC}, very close values are obtained for both multivariate modelling approaches M2 and M3.

An interesting aspect of the automatic group construction approach, which relies on the posterior correlation matrix of the linear predictor, is its inherent interpretability. For each testing point (area $i$, time $t$ and response $j$), we can examine its derived groups to identify which structured effect of the model contributes to a stronger correlation in the risk estimates. In the independent model (M1), the groups assigned to all the testing points consistently correspond to adjacent years. In the M-model (M2), while the temporally structured random effect remains as the dominant pattern in the model, there are some testing points where the group of highly correlated data points also includes spatial neighbours (areas that share a common border). Finally, the groups derived from the shared-component model (M3) reveal a strong correlation between both responses (mortality and incidence data) for the majority of areas in the year 2011. This finding suggest that the shared-component model would be the most suitable choice for predicting pancreatic cancer mortality based on incidence data.

\begin{table}[!ht]
\caption{Pancreatic cancer data: model selection criteria (DIC) and predictive performance measures (Logarithmic Score) with automatic groups construction for $m=3,5$, and $10$.}
\label{tab:CancerData_DIC}
\begin{center}
\begin{tabular}{lrrrr}
\toprule
& \multirow{2}{*}{DIC} & \multicolumn{3}{c}{Logarithmic Score} \\
& & $m=3$ & $m=5$ & $m=10$ \\[0.5ex]
\hline
M1 - Type I   & 26095.3 & 3.113 & 3.115 & 3.122 \\
M1 - Type II  & {\bf 26065.4} & {\bf 3.109} & {\bf 3.112} & {\bf 3.119} \\
M1 - Type III & 26120.9 & 3.115 & 3.118 & 3.125 \\
M1 - Type IV  & 26074.1 & 3.110 & 3.112 & 3.119 \\
\hline
M2 - Type I   & 26000.0 & 3.099 & 3.101 & 3.104 \\
M2 - Type II  & {\bf 25961.7} & {\bf 3.094} & {\bf 3.093} & {\bf 3.094} \\
M2 - Type III & 26018.8 & 3.101 & 3.102 & 3.106 \\
M2 - Type IV  & 25973.5 & 3.095 & 3.095 & 3.096 \\
\hline
M3 - Type I   & 25995.0 & 3.098 & 3.099 & 3.106 \\
M3 - Type II  & {\bf 25953.9} & {\bf 3.092} & {\bf 3.092} & {\bf 3.099} \\
M3 - Type III & 26012.1 & 3.099 & 3.100 & 3.107 \\
M3 - Type IV  & 25964.7 & 3.093 & 3.093 & 3.100 \\
\bottomrule
\end{tabular}
\end{center}
\end{table}

\subsection{Spatial Compositional Data: the case of \textit{Arabidopsis thaliana}} \label{sec:arabidopsis}

Compositional Data Analysis has gained popularity for comprehending processes wherein values correspond to $D$ distinct categories, with their sum constituting a constant. These values typically represent proportions (or percentages) summing up to one (or 100). The data stemming from such processes are commonly referred to as Compositional Data (CoDa). Without loss of generality, we assume the constant to be one. A substantial body of literature exists, delving into the application of CoDa analysis across various domains, including Ecology \citep{kobal2017, douma2019}, Geology \citep{buccianti2014, engle2014}, Genomics \citep{tsilimigras2016,shi2016,washburne2017,creus2022}, Environmental Sciences \citep{aguilera2021, mota2022} or Medicine \citep{dumuid2018, fairclough2018}.


Recently, two approaches have been introduced to incorporate the CoDa analysis within the framework of latent Gaussian models. One involves utilizing Dirichlet regression \citep{martinez-minaya2023dirinla}, while the other proposes the Logistic Normal Dirichlet Model (LNDM) \citep{martinez-minaya2023INLAcomp}, which mainly uses logistic-normal distribution with Dirichlet covariance through the additive log-ratio transformation as likelihood promoting interpretation as log-ratios. In this example, we focus on the latter, illustrating the utility of an automatic group construction procedure for LGOCV in a spatial CoDa problem: the case of \textit{Arabidopsis thaliana}.

\subsubsection{The data}
We conduct our study using a dataset consisting of $N=301$ accessions of the annual plant \textit{Arabidopsis thaliana} located on the Iberian Peninsula. For each accession, the genetic cluster (GC) membership proportions for the four genetic clusters (GC1, GC2, GC3, GC4) are available as identified in \cite{martinez-minaya2019}. These probabilities, shown in \autoref{fig:coda_arabidopsis_raw_alr}, sum up to one. Our main interest lies in estimating the membership probability, which in this specific context can be interpreted as the habitat suitability for each genetic cluster. To achieve this, we employ LNDMs incorporating climate-related covariates and spatially structured effects within the linear predictor. Specifically, we employ two bioclimatic variables to define the climatic aspect: annual mean temperature (BIO1) and annual precipitation (BIO12). The dataset is available at \url{http://doi.org/10.5281/zenodo.2552025} \citep{martinez-minaya2019}. Prior to analysis, climate covariates are standardized.

\subsubsection{The models}
\begin{figure}[!ht]
    \includegraphics[width=\textwidth]{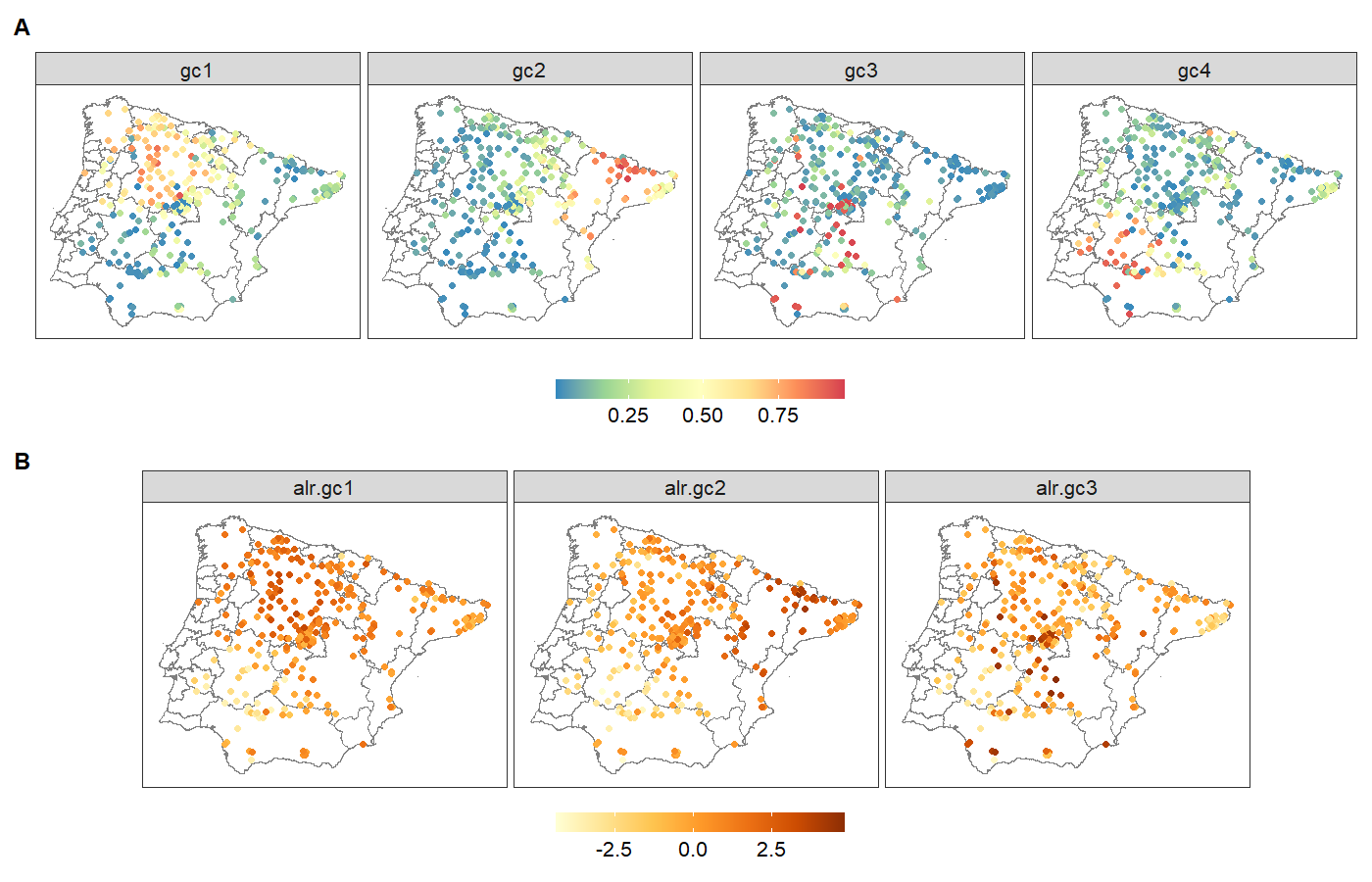}
    \caption{CoDa of 301 accessions of the annual plant \textit{Arabidopsis thaliana} located on the Iberian Peninsula. \textbf{A}: Membership proportion to each of the 4 genetic clusters extracted from \cite{martinez-minaya2019}. \textbf{B}: $alr$-coordinates generated from the $alr$ transformation using GC4 as reference.}
    \label{fig:coda_arabidopsis_raw_alr}
\end{figure}

As mentioned earlier, four categories are employed in this problem: GC1, GC2, GC3 and GC4. So, we deal with proportions in the simplex $\mathbb{S}^4$. To construct the LNDM, the additive log-ratio transformation ($alr$) is required, and we select GC4 as the reference category because it has the lowest variance in its logarithm. We are thus dealing with a three dimensional normal distribution with Dirichlet covariance, $\mathcal{ND}(\boldsymbol{\mu}, \boldsymbol{\Sigma})$. The data, as well as the transformed data, are displayed in \autoref{fig:coda_arabidopsis_raw_alr}.

The models employed to address this issue present the subsequent structure:
\begin{eqnarray}
\begin{array}{rcl}
alr(\boldsymbol{Y} ) & \sim & \mathcal{ND}((\boldsymbol{\mu}^{(1)}, \ldots, \boldsymbol{\mu}^{(3)}), \boldsymbol{\Sigma}), \\[1.ex]
\boldsymbol{\mu}^{(d)} & = & \boldsymbol{X} \boldsymbol{\beta}^{(d)} + \boldsymbol{\omega}^{(d)} \,, \ d = 1, \ldots, 3,
\end{array}
\label{eq:model_arabidopsis}
\end{eqnarray}
where $alr(\boldsymbol{Y})$ represents the $alr$ transformation applied to the compositional variable. From now on, we will refer to the transformed variable as $alr$-coordinates, and the vector $\boldsymbol{\mu}^{(d)} = (\mu^{(d)}_{1}, \ldots, \mu^{(d)}_{301})^{'}$ is the mean of the $d$-th $alr$-coordinate. The design matrix $\boldsymbol{X}$ of dimension $301 \times 3$ consists of ones in its first column and climate covariate values in the remaining columns. The spatial random effect for each $d$-th $alr$-coordinate is denoted by $\boldsymbol{\omega}^{(d)}$, characterized by a Mat\'ern covariance structure. Specifically, $\boldsymbol{\omega}^{(d)}$ follows a normal distribution $\mathcal{N}(0, \boldsymbol{Q}^{-1}(\sigma_{\boldsymbol{\omega}}, \phi))$, where $\sigma_{\boldsymbol{\omega}}$ represents the standard deviation of the spatial effect, and $\phi$ its range. The parameter vector $\boldsymbol{\beta}=(\beta^{(d)}_{0}, \beta^{(d)}_{1}, \beta^{(d)}_{2})^{'}$ corresponds to fixed effects. The latent field comprises both the fixed effects parameters and realizations of the random field. This configuration can be succinctly presented as:
\begin{equation*}
\boldsymbol{\mathcal{X}} = \{\boldsymbol{\beta}^{(d)}, \boldsymbol{\omega}^{(d)}  : d = 1, \ldots, 3\}.
\end{equation*}
In contrast, the vector $\boldsymbol{\theta}_1 = \{\sigma_d^2, \gamma : d = 1, \ldots, 3 \}$ encompasses hyperparameters linked to the likelihood. On the other hand, $\boldsymbol{\theta}_2 = \{\sigma_{\boldsymbol{\omega}}, \phi\} $ comprises hyperparameters pertaining to the spatial random effect. These combined elements collectively form the set of hyperparameters. Gaussian prior distributions are assumed for the fixed effects, while PC-priors are employed for the hyperparameters, following \cite{simpson_penalising_2017}.
Based on the model structure defined in Equation \eqref{eq:model_arabidopsis} where a spatial effect is included, we focus on the eight structures presented in \cite{martinez-minaya2023INLAcomp}:
\begin{itemize}
   \item Type I: share the same parameters for fixed effects, and do not include spatial random effects.
    \item Type II: have different parameters for fixed effects, and do not include spatial random effects.
    \item Type III: share the same parameters for fixed effects, and share the same spatial effect.
    \item Type IV: different parameters for fixed effects, and share the same spatial effect.
    \item Type V: share the same parameters for fixed effects, and the spatial effects between linear predictors are proportional. 
    \item Type VI: different parameters for fixed effects, and the spatial effects between linear predictors are proportional. 
    \item Type VII: share the same parameters for fixed effects, and different realisations of the spatial effect for each linear predictor. 
    \item Type VIII: different parameters for fixed effects, and different realisations of the spatial effect for each linear predictor. 
\end{itemize}
Note that in Type V and Type VI structures, realisations of the spatial field are the same, but a proportionality hyperparameter is added in two of the three linear predictors, denoted as $\alpha^{(1)}$ and $\alpha^{(2)}$. On the other hand, in Type VII and Type VIII structures, although realisations of random effects are different, they share the same hyperparameters. \autoref{tab:coda_s-arabidopsis} provides an overview of the distinct structures, their associated latent fields, and the accompanying set of hyperparameters. 

\begin{table}[!ht]
\caption{Different structures included in the model in an additive way with their corresponding latent field and the set hyperparameters to be estimated.}
\label{tab:coda_s-arabidopsis}
\begin{center}
\vspace{-0.5cm}
\resizebox{\textwidth}{!}{
\begin{tabular}{llll}
\toprule
\em Models &\em Predictor &\em Latent Field ($\boldsymbol{\mathcal{X}}$) &\em Hyperparameters ($\boldsymbol{\theta}$) \\
\hline \\ [-0.3cm]
Type I & $\boldsymbol{X} \boldsymbol{\beta}$ & $\{\beta_0, \beta_1, \beta_2\}$ & $
\{\sigma_d^2, \gamma\}$\\[0.1cm]
Type II & $\boldsymbol{X} \boldsymbol{\beta}^{(d)}$ & $\{\beta_0^{(d)}, \beta_1^{(d)}, \beta_2^{(d)}\}$ & $
\{\sigma_d^2, \gamma\}$\\[0.1cm]
\hline \\ [-0.3cm]
Type III & $\boldsymbol{X} \boldsymbol{\beta} + \boldsymbol{\omega}$ &  $\{\beta_0, \beta_1, \beta_2, \omega_{1}, \ldots, \omega_{N}\}$ & $
\{\sigma_d^2, \gamma, \sigma_{\boldsymbol{\omega}}, \phi\}$\\[0.1cm]
Type IV & $\boldsymbol{X} \boldsymbol{\beta}^{(d)} + \boldsymbol{\omega}$ &  $\{\beta_0^{(d)}, \beta_1^{(d)}, \beta_2^{(d)}, \omega_{1}, \ldots, \omega_{N}\}$ & $
\{\sigma_d^2, \gamma, \sigma_{\boldsymbol{\omega}}, \phi\}$\\[0.1cm]
\hline \\ [-0.3cm]
Type V & $\boldsymbol{X} \boldsymbol{\beta} + \boldsymbol{\omega}^{*(d)}$ & $\{\beta_0, \beta_1, \beta_2, \omega_{1}, \ldots, \omega_{N}\}$ & $
\{\sigma_d^2, \gamma, \sigma_{\boldsymbol{\omega}}, \phi, \alpha^{(1)}, \alpha^{(2)} \}$\\ [0.1cm]
Type VI & $\boldsymbol{X} \boldsymbol{\beta}^{(d)} + \boldsymbol{\omega}^{*(d)}$ & $\{\beta_0^{(d)}, \beta_1^{(d)}, \beta_2^{(d)}, \omega_{1}, \ldots, \omega_{N}\}$ & $
\{\sigma_d^2, \gamma, \sigma_{\boldsymbol{\omega}}, \phi, \alpha^{(1)}, \alpha^{(2)} \}$\\[0.1cm]
\hline \\ [-0.3cm]
Type VII & $\boldsymbol{X} \boldsymbol{\beta} + \boldsymbol{\omega}^{(d)}$ & $\{\beta_0, \beta_1, \beta_2, \omega_{1}^{(d)}, \ldots, \omega_{N}^{(d)}\}$ & $
\{\sigma_d^2, \gamma, \sigma_{\boldsymbol{\omega}}, \phi\}$\\[0.1cm]
Type VIII & $\boldsymbol{X} \boldsymbol{\beta}^{(d)} + \boldsymbol{\omega}^{(d)}$ & $\{\beta_0^{(d)}, \beta_1^{(d)}, \beta_2^{(d)}, \omega_{1}^{(d)}, \ldots, \omega_{N}^{(d)}\}$ & $
\{\sigma_d^2, \gamma, \sigma_{\boldsymbol{\omega}}, \phi \}$\\
\bottomrule
\end{tabular}}
\end{center}
\end{table}

\subsubsection{LGOCV in Compositional Data}
As highlighted in \cite{martinez-minaya2023INLAcomp}, excluding a category from a CoDa point during cross-validation procedures may not be practical due to the inherent constraint imposed by CoDa, where the sum of its components must equal one. 
This implies that the remaining categories hold valuable information about the category we intend to exclude. Consequently, the remaining log-ratio coordinates offer insights into the category we have removed during cross-validation. In this manner, the concept of `friendship' emerges. Accordingly, we can assert that the first $alr$-coordinate of $i$-th data point is associated with the second $alr$-coordinate of $i$-th data point, thereby contributing pertinent information.

Hence, to perform cross-validation for the $i$-th data point and the $d$-th $alr$-coordinate, it becomes necessary to exclude the values associated with all $alr$-coordinates for that data point and for the corresponding group. To do this, the group $I_i^{(d)}$ must be constructed. If it is determined that the $i$-th data point in its $d$-th $alr$-coordinate is associated with the data point indexed as $i + 1$ in its $d$-th $alr$-coordinate, the data point $i+1$ will also be included in the set $I_i^{(d)}$ in all of its $alr$-coordinates. It should be noted that $I_{i}^{(d)}$ could be different from $I_{i}^{(d^{*})}$, being $d \neq d^*$, as even though we are referring to the same compositional point $i$, we are modelling each $alr$-coordinate. So we need to compute LS measures for CoDa in the following way:
\begin{eqnarray*}
    \mbox{LS} & = & - \frac{1}{N \cdot (D-1)} \sum\limits_{d = 1}^{D-1} \sum \limits_{i = 1}^N \log \left(\pi(\text{alr}(\boldsymbol{y})_i^{(d)} \mid \text{alr}(\boldsymbol{y})_{-I_{i}^{(d)}}^{\bullet})\right),
    %
\end{eqnarray*}
where $alr(\boldsymbol{y})_i^{(d)}$ is the observed vector for the $i$-th data point and the $d$-th $alr$-coordinate, and $alr(\boldsymbol{y})_{-i}^{\bullet}$ represents the observed data in $alr$-coordinates excluding the $I_i^{(d)}$ data points with its corresponding $D-1$ $alr$-coordinates.

\subsubsection{Results}
\autoref{tab:coda_s-arabidopsis-DIC} presents a comparison of various models used for estimating the membership proportions to the different genetic clusters of \textit{Arabidopsis thaliana}. As in the previous section, the evaluation is based on model selection criteria and predictive performance measures. To compute comparable LS measures from the posterior predictive density functions, we use the groups derived from Type VIII model. We use values of $m=3,5$ and $10$ for constructing the groups and compute performance measures based on LGOCV. Although, Type VII model seems to have a good predictive performance, DIC/WAIC and predictive measures based on LGOCV approaches consistently suggest that Type VIII model is the most suitable choice.

\begin{table}[!ht]
\caption{\textit{Arabidopsis thaliana} data: model selection criteria (DIC) and predictive performance measures (Logarithmic Score) with automatic groups construction for $m=3,5$, and $10$.}
\label{tab:coda_s-arabidopsis-DIC}
\begin{center}
\begin{tabular}{lrrrr}
\toprule
\multirow{2}{*}{Type} & \multirow{2}{*}{DIC} & \multicolumn{3}{c}{Logarithmic Score} \\
& & $m=3$ & $m=5$ & $m=10$ \\[0.1cm]
\hline
& & & & \\ [-0.5cm]
   I & 3353.403 & 1.894 & 1.897 & 1.904 \\[0.1cm]
  II & 3294.855 & 1.869 & 1.873 & 1.883 \\[0.1cm]
\hline
& & & & \\ [-0.5cm]
 III & 3202.082 & 1.784 & 1.780 & 1.795 \\[0.1cm]
  IV & 3145.198 & 1.757 & 1.756 & 1.778 \\[0.1cm]
\hline
& & & & \\ [-0.5cm]
   V & 3059.359 & 1.470 & 1.579 & 1.634 \\[0.1cm]
  VI & 3002.910 & 1.396 & 1.526 & 1.588 \\[0.1cm]
\hline
& & & & \\ [-0.5cm]
 VII & 2754.844 & 1.427 & \bf{1.513} & \bf{1.567} \\[0.1cm]
VIII & \bf{2741.440} & \bf{1.415} & \bf{1.513} & \bf{1.568} \\[0.1cm]
  \hline
\end{tabular}
\end{center}
\end{table}

\begin{figure}[!ht]
    \includegraphics[width=\textwidth]{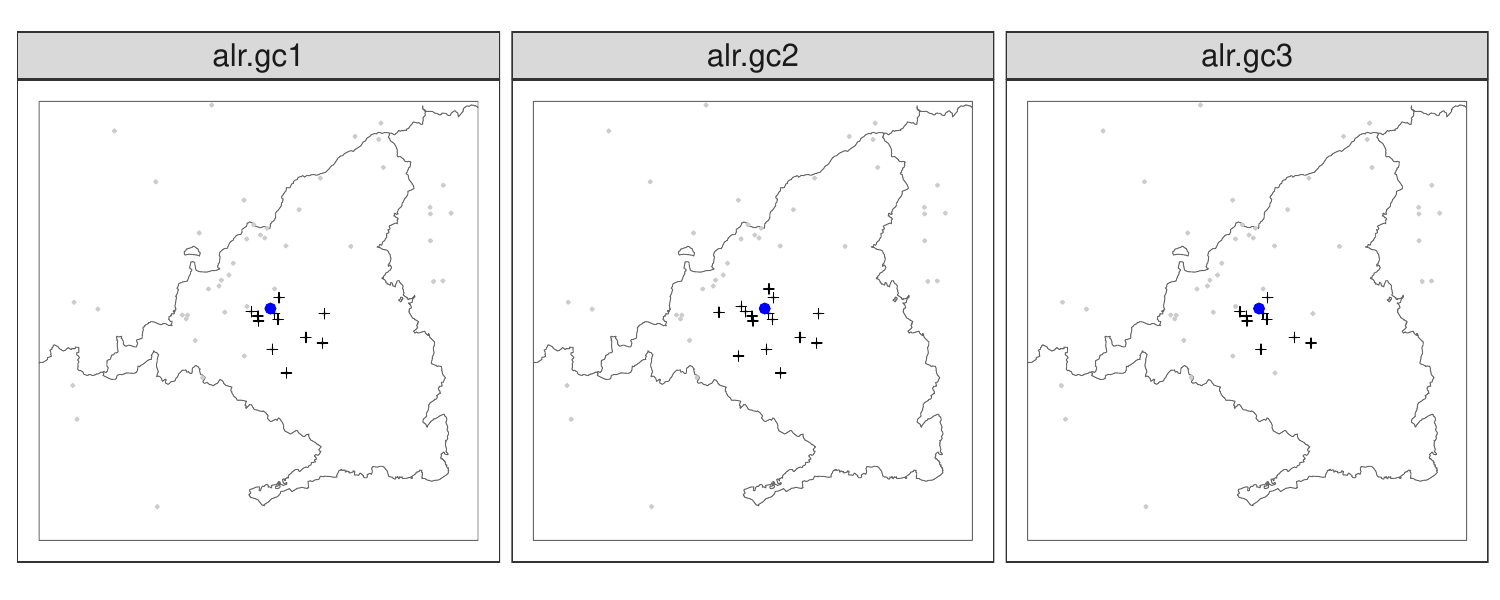}
    \vspace{-0.8cm}
    \caption{Automatically constructed groups (with $m=10$) under Type VIII model for data point $i=88$ located in the region of Madrid. The blue circle represents the focal point, while the crosses indicate the points that belong to the constructed group $I_{88}^{(d)}$, $d = 1, \ldots, 3$. \label{fig:coda_groups_typeVIII}}
\end{figure}

In \autoref{fig:coda_groups_typeVIII}, we depict automatic groups formed from the posterior correlations in the Type VIII model. Specifically, this representation pertains to the three $alr$-coordinates associated with the $88$-th data point when $m=10$. By employing the `friendship' feature, we made the assumption that these three observations share a bond, meaning that if we intend to create groups for the first $alr$-coordinate of observation 88, we must not only include that $alr$-coordinate of observation $i=88$ in $I_{88}^{(1)}$, but also the second and third $alr$-coordinates for that data point. As shown in \autoref{fig:coda_groups_typeVIII}, it becomes evident that the groups established for data point $i=88$ differ for each $alr$-coordinate, i.e., $I_{88}^{(1)} \neq I_{88}^{(2)} \neq I_{88}^{(3)}$. This observation underscores not only our assumption of distinct realizations of the spatial effect for each $alr$-coordinate but also varying effects of the climatic variables. This is due to the fact that posterior correlations consider all components of the linear predictor.

\subsection{Space-time models for the United Kingdom wind speed data}\label{sec:spacetime}

\begin{figure}[!htb]
    \begin{center}
    \includegraphics[width=0.5\textwidth]{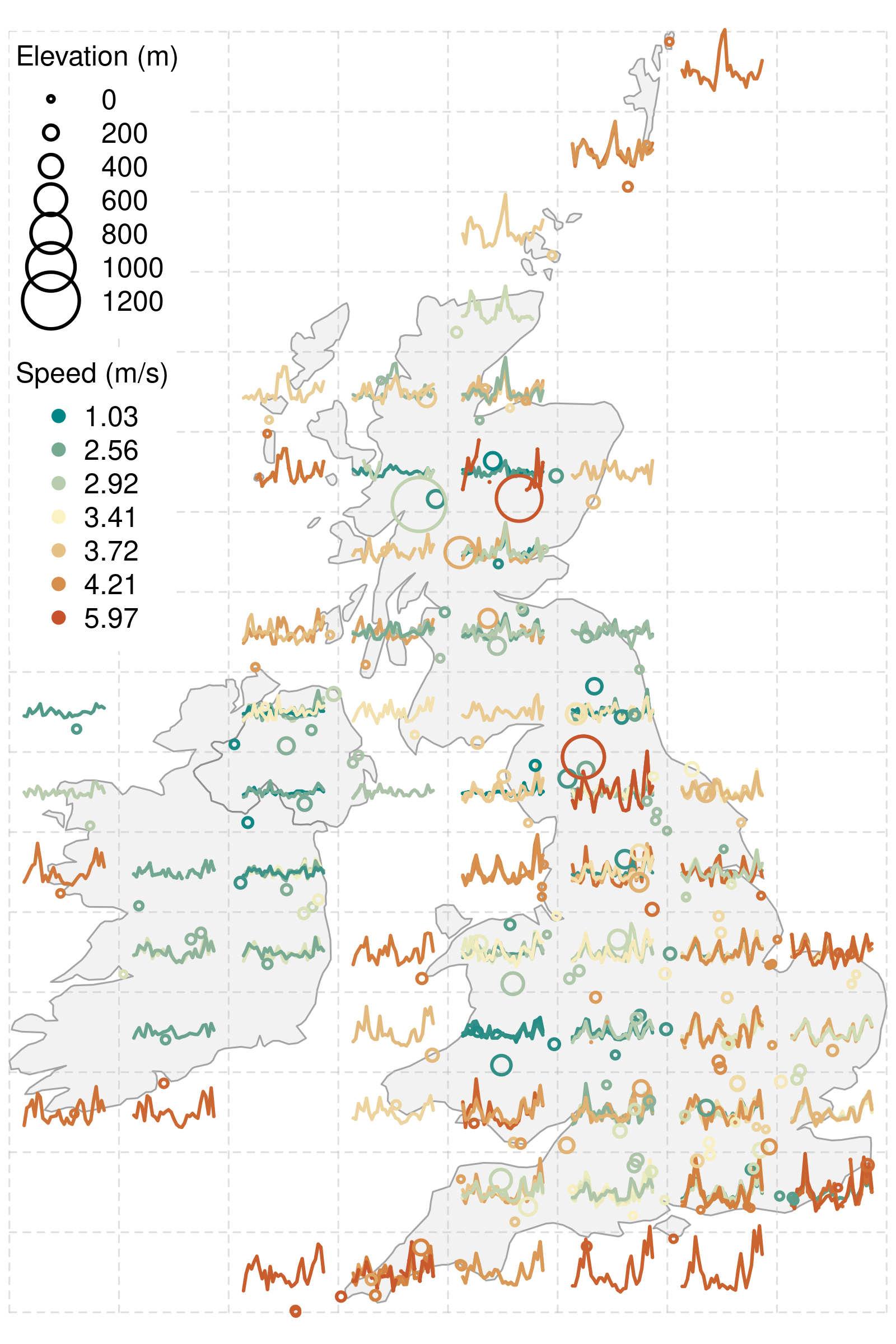}
    \end{center}
    \caption{Map displaying the station locations with the data presented as grouped time series (respect to the spatial grid overlayed in dashed gray lines) over the period of July 2021.}
    \label{fig:wdatastations}
\end{figure}

In what follows, we describe how to use the procedure outlined in Section~\ref{sec:LGOCV} to perform cross-validation on space-time models for wind speed data in the United Kingdom (UK). We consider a dataset from a regularly updated 20th and 21st-century database from the European Climate Assessment \& Dataset (ECA\&D) project \citep{klein2002daily}. We select daily wind speed measurements, in meters per second (m/s), from July 2021 at 189 stations in Ireland and the UK. The locations of the stations are presented as coloured dots in \autoref{fig:wdatastations}. The observations at each location are displayed with different colors in this figure as grouped time series over July 2021. The grouped time series are constructed by computing the average daily wind speed over the stations located within each spatial grid (outlined by dashed gray lines).

\subsubsection{The models} \label{sec:models:spacetime}

We compare four models with different characterizations of the space-time dynamics in the UK wind speed data. Consider the process $Y(\mathbf{s}, t)$, which represents the wind speed at location $\mathbf{s}$ and at time $t$. We model observations $y(\mathbf{s}_i, t_i)$ from this process collected at site $\mathbf{s}_i$ and time $t_i$ with a Gaussian distribution N($\mu(\mathbf{s}_i, t_i)$, $\sigma^2_e$), where $\sigma^{2}_{e}$ is a  common variance parameter and the mean function is given by
\begin{equation}
\mu(\mathbf{s}, t) = \beta_{0} + \beta_{1} x(\mathbf{s}) + r(\mathbf{s}) + u(\mathbf{s}, t),
\label{eq:latent-Y}
\end{equation}
where $\beta_{0}$ is the intercept, which is constant in space and time, $x$ is the elevation (in meters), and $\beta_1$ represents the associated regression coefficient. The $r(.)$ term is a spatial field, which accommodates regional persistent spatial variations and is modeled with a Mat\'ern covariance using the stochastic partial differential equation (SPDE) approach described in \cite{lindgren2011explicit} with smoothness parameter fixed to $\alpha=2$, range $\rho_r$, and marginal standard error $\sigma_r$, which need to be estimated. Space-time interactions are captured through the space-time random field $u(.)$, defined using the models proposed in~\cite{lindgren2023diffusion}. These models are defined as solutions to a diffusion-based family of equations that induces physically realistic processes, such as wind. Specifically, it is the solution to the following SPDE
\begin{equation}
\gamma_e(\gamma_s^2 - \Delta)^{\alpha_e/2}\left( \gamma_t \frac{\partial}{\partial t} + (\gamma_s^2 - \Delta)^{\alpha_s} \right)^{\alpha_t} \ u(s,t) = \mathcal{W}(s,t).
\label{eq:demf_spde}
\end{equation}
The power parameters $\alpha_t$, $\alpha_s$ and $\alpha_e$ in Equation~\eqref{eq:demf_spde} specify the smoothness of $u(.)$, whereas $\gamma_s$, $\gamma_t$, and $\gamma_e$ govern the marginal properties and are related to dynamics of the process. We consider here four cases of the models proposed in \cite{lindgren2023diffusion}:
Model $M_A$ (with $\alpha_t = 1$, $\alpha_s=0$, and $\alpha_e=2$);
Model $M_B$ (with $\alpha_t = 1$, $\alpha_s=2$, and $\alpha_e=1$);
Model $M_C$ (with $\alpha_t = 2$, $\alpha_s=0$, and $\alpha_e=2$);
and Model $M_D$ (with $\alpha_t = 2$, $\alpha_s=2$, and $\alpha_e=0$).
When $\alpha_s=0$ (Models $M_A$ and $M_C$), the space-time SPDE in Equation~\eqref{eq:demf_spde} generates a precision matrix that can be factorized into two precision matrices, one for the purely spatial model and another for the purely temporal model. While separability is broadly used due to a substantial reduction of the computation time when working with large data sets with complex dependencies, it is usually too simplistic to model natural phenomena. Due to recent developments, with the SPDE approach both, separable or non-separable, have similar computation time.

For fitting the models, we considered a flat prior for $\beta_0$ and a Gaussian with zero mean and variance $0.01$ for $\beta_1$.
The penalized complexity prior proposed in \cite{fuglstad2018pc},
are used for the spatial and temporal range parameters of $r$ and $u$, as well as for 
all the variance parameters $\sigma^2_e$, $\sigma^2_r$ (marginal variance of $r$), and $\sigma^2_u$ (marginal variance of $u$).

\subsubsection{Results}

\begin{figure}[!htb]
    \begin{center}
    \includegraphics[width=0.75\textwidth]{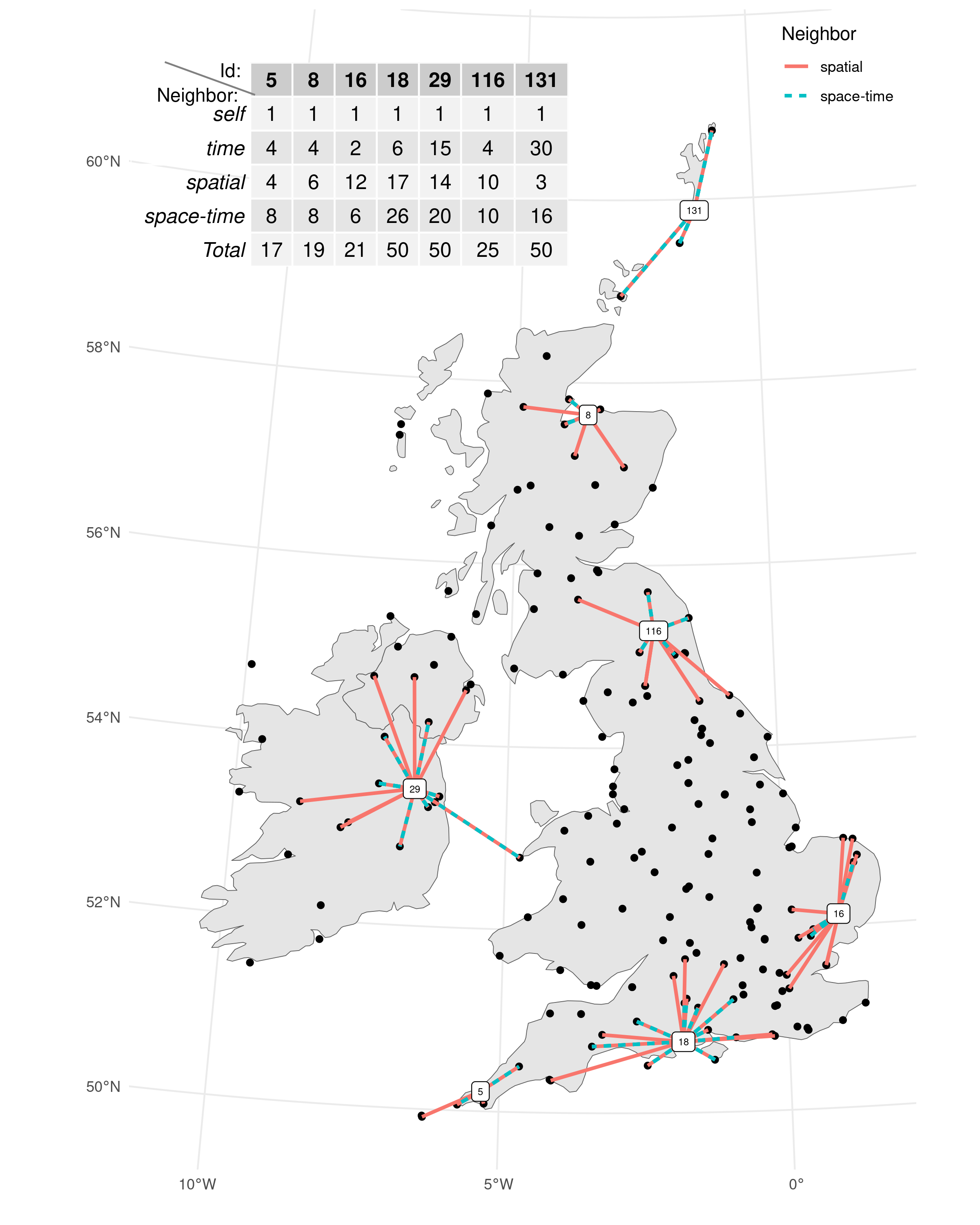}
    \end{center}
    \caption{Location of the stations and examples of seven observations at time 16 with their respective automatic selected groups from the LGOCV method.}
    \label{fig:stgroups}
\end{figure}

We use the methods described in Section~\ref{sec:LGOCV} to perform LGOCV on the four models presented above. For all the models, we set the groups to those generated from model $M_A$. \autoref{fig:stgroups} illustrates the groups for $m=10$, at time $1$ for seven locations from different parts of the spatial domain.
Even though $m$ is fixed to 10, the total number of observations in each group is always larger than $10$ (see the last row of the summary table on the top left of this figure). This is due to several observation pairs having very similar correlation values.
For instance, the group for station `id' = 5 is of size 17 and for `id' = 131 it is of size 50, which is the number we set as the maximum group size.
In the top left table, for each station, the removed neighbours are classified into \textit{self} (same location and time), \textit{time} (same location but different times), \textit{spatial} (same time but different locations), or \textit{space-time} (different location and different time). Stations located in isolated places such as `id' = 131 tend to select more neighbours in time. Indeed, only three spatial neighbors were selected and the entire time series of this station was left out, that is fifteen observations before and fifteen after the selected time. On the other hand, stations that have several several neighbors close by such as `id' =  18 tend to mostly include spatial and space-time neighbours.

Finally, \autoref{tab:stlgpo} shows the DIC for the different models and the LS values for $m=3$, $m=5$, and $m=10$. Model $M_B$ provides the best fit with the lowest DIC. This model also performed the best for prediction purposes, as shown by the LS values for all level sets. Model $M_D$ has the second lowest DIC but the worst LS, indicating that it is too smooth to predict the highly variable wind trajectories (see \autoref{fig:wdatastations}).



\begin{table}[!ht]
\caption{wind speed data: model selection criteria (DIC) and predictive performance measures (Logarithmic Score) with automatic groups construction for $m=3$, $m=5$, and $m=10$.}
\label{tab:stlgpo}
\begin{center}
\begin{tabular}{crrrr}
\hline
\multirow{2}{*}{Model for $u$} & \multirow{2}{*}{DIC} & \multicolumn{3}{c}{Logarithmic Score} \\
& & m=3 & m=5 & m=10 \\
\hline
$M_A$ & 3499.78 & 6202.79 & 6692.26 & 7461.16 \\
$M_B$ & \textbf{1622.16} & \textbf{6054.93} & \textbf{6563.31} & \textbf{7413.71} \\
$M_C$ & 3433.40 & 6214.94 & 6678.89 & 7435.29 \\
$M_D$ & 2078.38 & 7123.03 & 7587.39 & 8342.42 \\
\hline
\end{tabular}
\end{center}
\end{table}

\section{Discussion} \label{sec:Discussion}

Cross-validation techniques are at the core of any prediction task as they reveal the adequacy of a statistical model to unobserved data. Among the several cross-validation methods developed over the years, LOOCV stands out as the prevailing choice in numerous practical applications due to its simplicity and straightforward implementation. Although widely used, this method is only optimal under the assumption of independent and identically distributed (iid) data, an assumption rarely verified in many practical applications. In contrast, the newly developed LGOCV method handles non iid data with multiple correlation structures by automatically excluding a determined set of data points from the training set, expanding beyond the practice of solely excluding individual observations. The R-INLA package provides a convenient implementation of LGOCV for latent Gaussian models that is highly efficient and readily accessible.

In this paper, we showed that the automatic group construction procedure for LGOCV provides a robust measure of predictive performance in many settings where dependencies in space and/or time are present. Our simulation studies and real data analysis underscore that a common practice in modelling dependent data often leads to erroneous model selection and interpretations. In contrast, the LGOCV method, integrated into the R-INLA package, stands as an improved alternative to LOOCV.
In the model selection process, the LGOCV framework should be the primary consideration, especially when evaluating the predictive extrapolation performance of models.

\section*{Acknowledgments}
This research has been supported by projects PID2020-115882RB-I00 (JM-M) and PID2020-113125RB-I00/MCIN/AEI/10.13039/501100011033 (AA).
We would like to thank the valuable comments made by two anonymous reviewers that have contributed to clarify some aspects of this paper.

\bibliographystyle{apalike}
\bibliography{refs}   

\setcounter{section}{0} 
\renewcommand{\thesection}{\Alph{section}}

\setcounter{figure}{0}
\renewcommand\thefigure{\thesection\arabic{figure}}

\setcounter{table}{0}
\renewcommand\thetable{\thesection\arabic{table}}

\clearpage
\section{Appendix: simulation study}

In this section we summarize the results from an additional simulation study. The purpose of this study was to evaluate the ability of the automatic groups generated from different candidate models in the context of model selection. We considered three different scenarios, each one with three different models: A, B and C. Datasets were generated for each model, and the three models were fitted as a candidate model to the simulated data. The automatic group construction procedure for leave-group-out cross validation (LGOCV) based on the posterior distribution from each model fit was computed. Additionally, an evaluation set was defined as the union of these three automatically generated groups. The evaluation set with lowest Logarithmic Score (LS) value for each of the three fitted models was considered as a reference. Finally, the differences from the LS values and this reference plus one were summarized over $100$ simulations and visualized in log-scale using box plots. This was performed with three different level sets: $m=5$, $m=10$ and $m=30$. The key finding suggests that the automatic groups generated from the posterior of a reasonable model fitted to a dataset is the recommended approach to evaluate different candidate models.

\subsection*{Scenario 1}
We considered the map of the $n=100$ counties in the North Carolina state. We assumed a Poisson likelihood model, $y_i \sim$ Poisson (E$_i\lambda_i$), for $i=1,...,100$, where $E_i$ was simulated for each dataset from a Gamma($4$, $0.001$) distribution. We set three different models for the linear predictor, $\eta_i = \log(\lambda_i)$, as described in the following table

\begin{table}[!ht]
\centering
\begin{tabular}{|c|l|l|}
\hline
Model & Linear predictor & Model description \\
\hline
 A & $\alpha + \beta x_i$ & intercept + fixed effect \\
 B & $\alpha + s_i$ & intercept + spatial random effect \\
 C & $\alpha + \beta x_i + s_i$ & intercept + fixed effect + spatial random effect\\
\hline
\end{tabular}
\end{table}

The values of $x$ were drawn from a uniform distribution U$(0,1)$. The covariate values, $x_i$, were sampled from a continuous distribution giving the number of non duplicated elements in $\mathbf{\eta}$ equal the number of data points. This allows the automatic groups generated from the posterior distribution of the fitted model $A$ to be considered as one of the evaluation sets.

\begin{figure}[!ht]
    \begin{center}
    \vspace{-0.3cm}
    \includegraphics[width=0.95\textwidth]{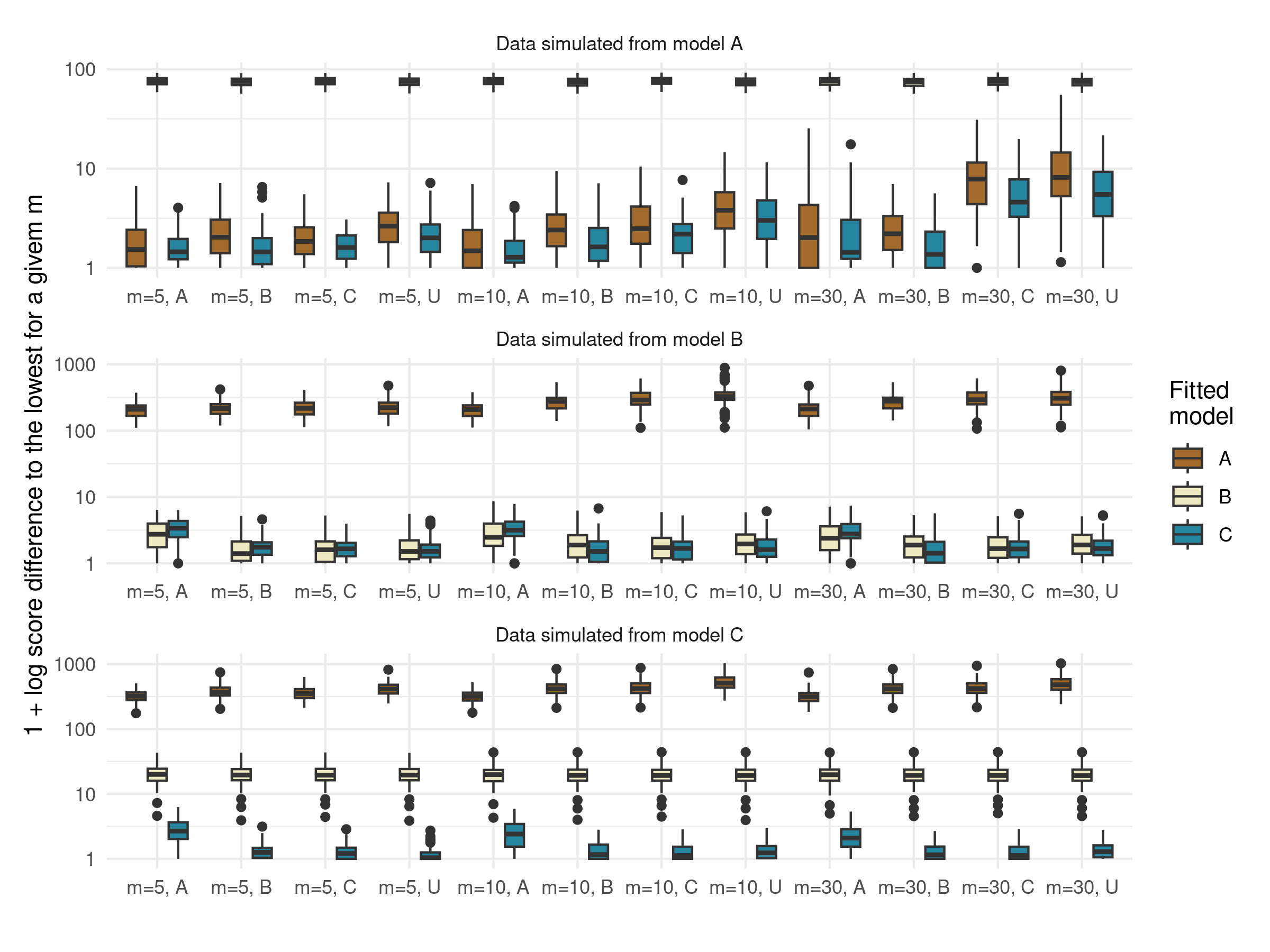}
    \vspace{-0.8cm}
    \end{center}
    \caption{Results from Scenario 1 by generating model and evaluation set.}
    \label{fig:cvsim1}
\end{figure}

The spatial effect $\mathbf{s}$ was sampled from a multivariate Gaussian distribution with zero mean vector and precision matrix $\tau(d_0 \mathbf{I} + \mathbf{R})$, where $\mathbf{I}$ is the identity matrix and
\[ \mathbf{R}_{ij} = \left\{
 \begin{array}{rl}
  n_i & \textrm{if }  i = j \\
  -1 & \textrm{if area } i \textrm{ is neighbor to area } j \\
  0 & \textrm{otherwise} \;.
 \end{array}
 \right.
\]
We used $d_0=1/3$ which was kept fixed and not estimated. For the precision parameter $\tau$, we used $\tau=1$ for the simulations and it was estimated considering a penalized complexity prior such that $P(1/\sqrt(\tau) = \sigma > 1) = 0.05$. The summarized results are shown in \autoref{fig:cvsim1}.

For the datasets simulated from model A, the lowest LS is obtained either for model A or model C. Model C performed well as it contains model A. Similarly, when the data was simulated from model B, either model B or model C have the lowest score. However, when the data was simulated from model C, the lowest score was always from model C. These conclusions hold for different values of $m$ and evaluation sets. Therefore, the automatic groups from any of the fitted models works fine for selecting the right model. In particular, even for the datasets simulated from models B or C, the automatic groups generated from model A works well.

\subsection*{Scenario 2}
This case considers a space-time setting with the same map of the previous scenario. Let consider $y_{it} \sim$ Poisson(E$_i\lambda_{ij}$), for $i$ = $1$, ..., $100$ and $t$ = $1$, ..., $30$, were $E_i$ was simulated for each dataset from a Gamma($4$, $0.001$) distribution. The three models for the linear predictor, $\eta_{it} = \log(\lambda_{it})$,
are defined as in the following table

\begin{table}[!ht]
\centering
\begin{tabular}{|c|l|l|}
\hline
Model & Linear predictor & Model description \\
\hline
 A & $\alpha + v_t + r_i$ & intercept + time effect + area effect \\
 B & $\alpha + v_t + r_i + u_{it} $ & model A + space-time unstructured effect \\
 C & $\alpha + v_t + r_i + s_{it} $ & model A + space-time structured effect \\
\hline
\end{tabular}
\end{table}

\begin{figure}[!ht]
    \begin{center}
    \vspace{-0.2cm}
    \includegraphics[width=0.95\textwidth]{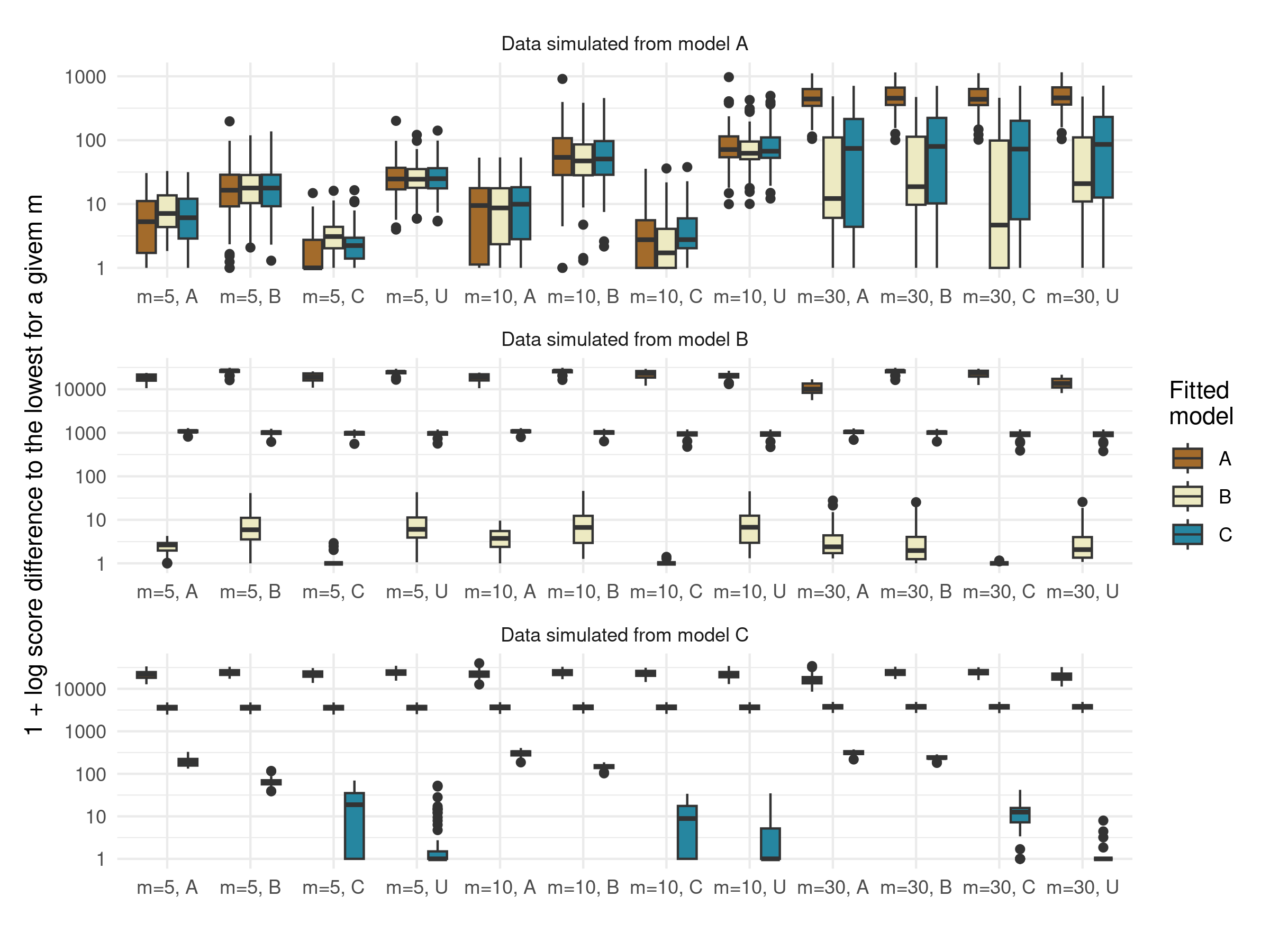}
    \vspace{-0.8cm}
    \end{center}
    \caption{Results from Scenario 1 by generating model and evaluation set.}
    \label{fig:cvsim2}
\end{figure}

The time effect $\mathbf{v}$ was simulated from a first order autoregressive model with autocorrelation parameter $\rho = 0.9$ (kept fixed during the estimation) and marginal variance $\sigma^2_v=1$, which was estimated considering a penalized complexity prior such that $P(\sigma > 1) = 0.05$. The spatial effect $\mathbf{r}$ was considered as in Scenario 1. The space-time effect $u_{it}$ was drawn from $N(0, 1)$ and fitted with the same prior as for $\sigma^2_v$. The space-time effect $d_{it}$ was drawn from the Type IV interaction model \citep{knorr2000}. The time and spatial structure matrices to build the space-time precision matrix for $\mathbf{d}$ were both scaled, as suggested in~\cite{sorbye2014scaling}. The precision parameter for $\mathbf{d}$ was also fitted with the same prior as for $\sigma^2_v$.
In \autoref{fig:cvsim2} we show the results for this scenario.

First, we will focus in the results for the simulated datasets from model A. Considering $m=5$ or $m=10$, either models A, B or C had the lowest LS values. This is not surprising, as both models B and C contain model A. Therefore, either models provide equally good predictions. The automatic groups generated with the posterior from the fitted model C tended to provide lower scores. This could be explained by considering that this model is the most structured one, thus providing the closest space-time neighbors as the evaluation set for each observation. With $m=30$ the fitted model A was the one providing the lowest LS values, so we conclude that for predicting a higher number of leave-group-out observations, the richer models that contains the true model performs better in this setting. For the datasets simulated from models B and C, the true generating models are selected as the best ones for all the evaluation sets. Therefore, using automatic groups built from the posterior of any of the fitted models worked in this scenario.

\subsection*{Scenario 3}
In Scenario 3, a continuous space-time domain with a linear model $y_i \sim N(\mu_i, \sigma^2_e)$ has been considered. We define three different models for the expected value $\mu_i$ of the $i$-th observation, taken at a location $(\mathbf{s}_i, t_i)$. Simulations were done considering the spatial domain as the square $[0,2]\times[0,2]$ and the time domain $[0, 20]$. The models are defined for $\mu(\mathbf{s}, t)$ as 

\begin{table}[!htb]
\centering
\begin{tabular}{|c|l|l|}
\hline
Model & Linear predictor & Model description \\
\hline
 A & $\alpha + v(t) + r(\mathbf{s})$ & int. + time effect + spatial effect \\
 B & $\alpha + v(t) + r(\mathbf{s}) + u_s(\mathbf{s},t)$ & A + separable space-time effect \\
 C & $\alpha + v(t) + r(\mathbf{s}) + u_u(\mathbf{s},t)$ & A + non-separable space-time effect \\
\hline
\end{tabular}
\end{table}

The model for $v(t)$ was set in a discrete time domain and handled the same way as in the previous scenario. The model for $r(\mathbf{s})$ is a Mat\'ern random field with smoothness equals $1$, range equals 2 and marginal variance equals 1. All these parameters were kept fixed during the estimation process. The models for $u_s(\mathbf{s},t)$ and $u_u(\mathbf{s},t)$ are described in \cite{lindgren2023diffusion}. What is separable or non-separable in these models is its covariance. For $u_s(\mathbf{s},t)$ we used the model with $\alpha_t=1$, $\alpha_s=0$ and $\alpha_e=2$, spatial range equals 1, time range equals 10 and variance equals 1. For $u_u(\mathbf{s},t)$ we used the model with $\alpha_t=1$, $\alpha_s=2$ and $\alpha_e=1$, spatial range equals 1, time range equals 20 and variance equals 1. In the estimation, we fixed the spatial range equals to the value used to simulate. Penalized complexity priors are considered for the time range and for the variance as suggested in \cite{lindgren2023diffusion}. Specifically, we set following probability statements: P(time range $< 2$) = $0.05$ for $u_s(\mathbf{s},t)$, P(time range $< 4$) = $0.05$ for $u_u(\mathbf{s},t)$,
and P(standard deviation $>1$) = $0.05$ in both cases.

\begin{figure}[!htb]
    \begin{center}
    \vspace{-0.2cm}
    \includegraphics[width=0.99\textwidth]{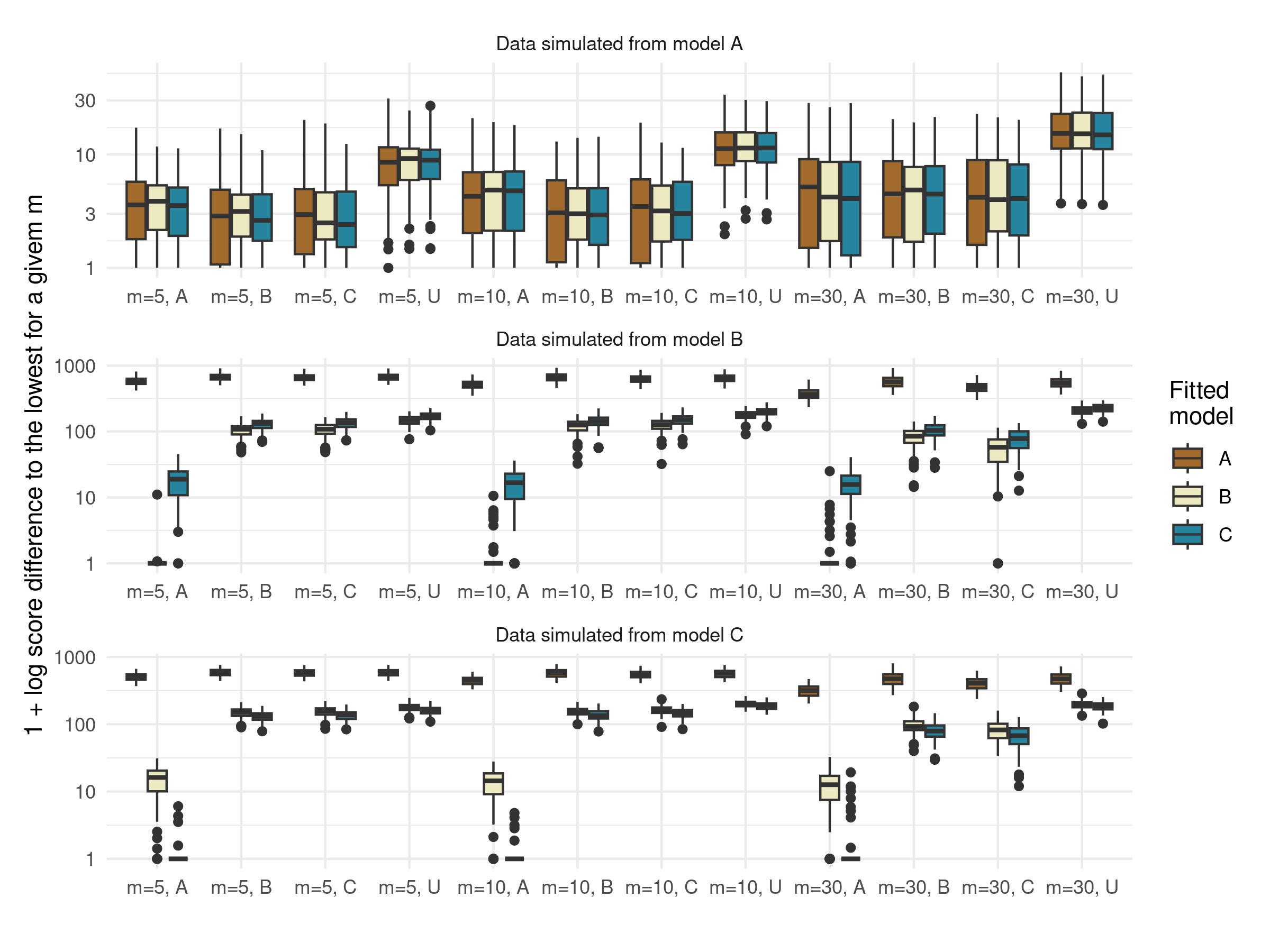}
    \vspace{-0.7cm}
    \end{center}
    \caption{Results from Scenario 3 by generating model and evaluation set.}
    \vspace{0.5cm}
    \label{fig:cvsim3}
\end{figure}

The results are shown in \autoref{fig:cvsim3}.For the datasets simulated from model A, the LS from different evaluation sets are similar for the different models fitted. For the other datasets, model A was never the best one, but the right model was not always selected according to the LS for all the evaluation sets. In general, the scores computed from automatic groups generated from the fitted model C are lower.
In particular, for datasets simulated from model B and $m=30$, consider fitting model B and using the automatic group from this fit to compute the LS. Then, fit model C and use the automatic group from this fit to compute the LS. According to the results in \autoref{fig:cvsim3}, this procedure tends to select model C more frequently. This fact illustrates that self-evaluation sets are not preferable. Instead, a common evaluation set is recommended to evaluate all the models fitted.

\pagebreak
\section{Appendix: additional figures}

\subsection*{Joint modelling of pancreatic cancer mortality and incidence data}

\begin{figure}[!ht]
    \begin{center}
        \includegraphics[width=\textwidth]{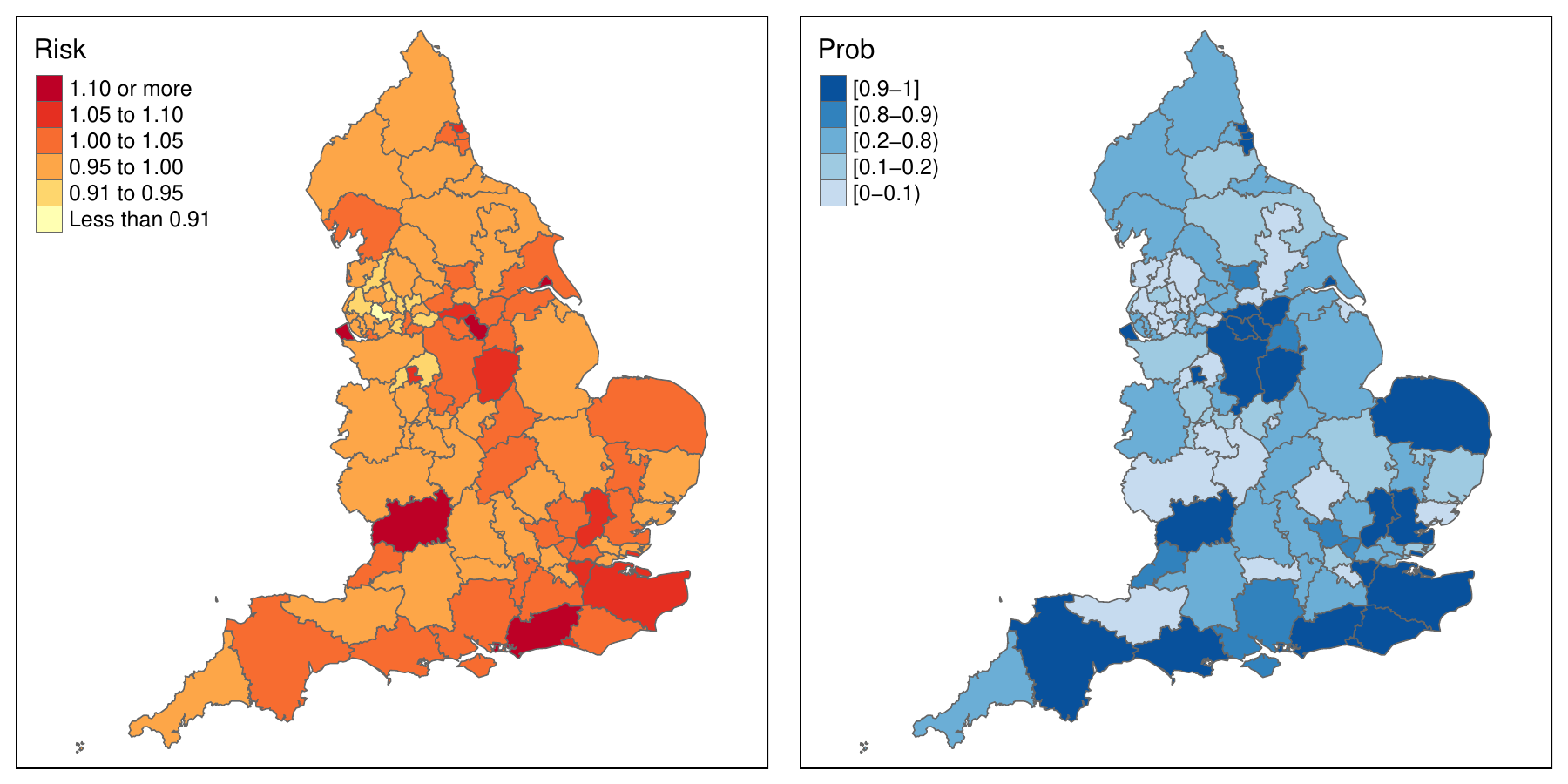}\\
        \includegraphics[width=0.7\textwidth]{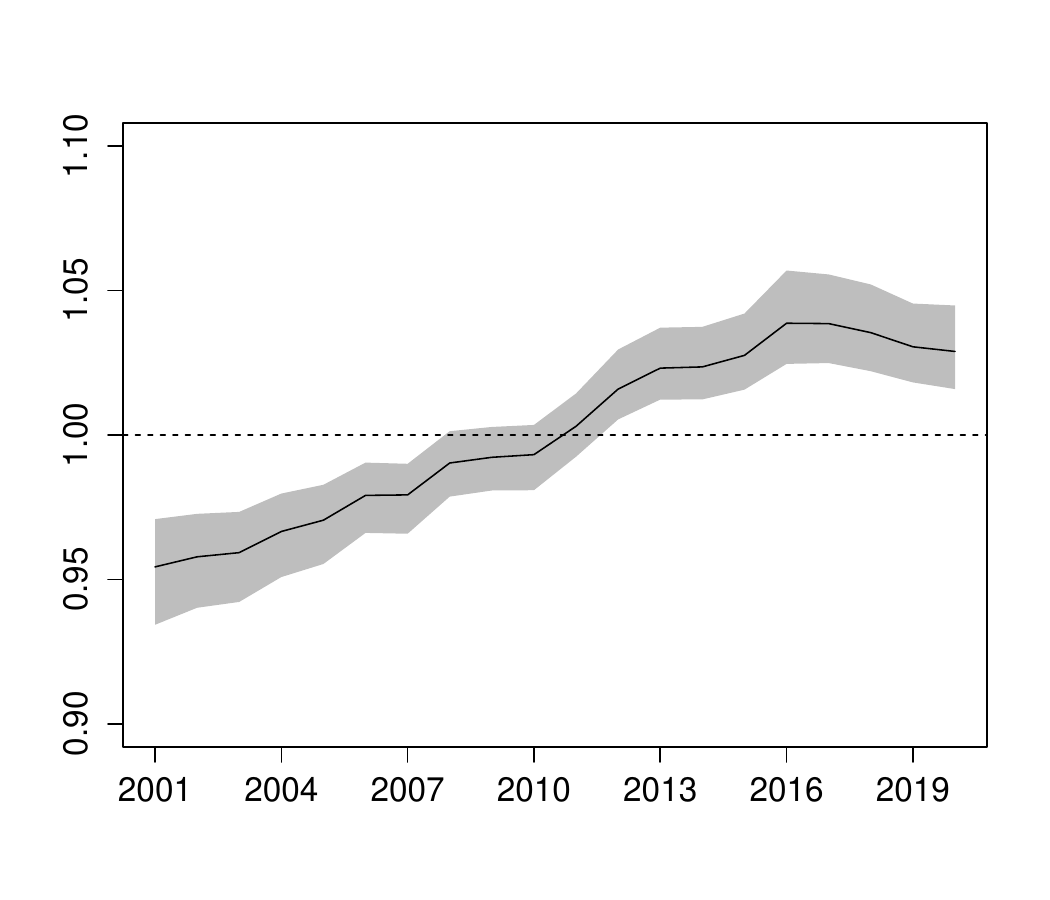}
        \vspace{-0.5cm}
    \end{center}
    \caption{Top: posterior median estimates $\exp(\phi_i)$ and posterior exceedence probabilities $P(\exp(\phi_i)>1 | \boldsymbol{y})$ for the shared spatial random effect (M3). Bottom: Posterior median estimates and 95\% credible intervals of the temporal shared random effect (M3 - TypeII). \label{fig:CancerData_Patterns}}
\end{figure}

\begin{figure}[!ht]
    \begin{center}
        \vspace{-1cm}
        \includegraphics[width=0.9\textwidth]{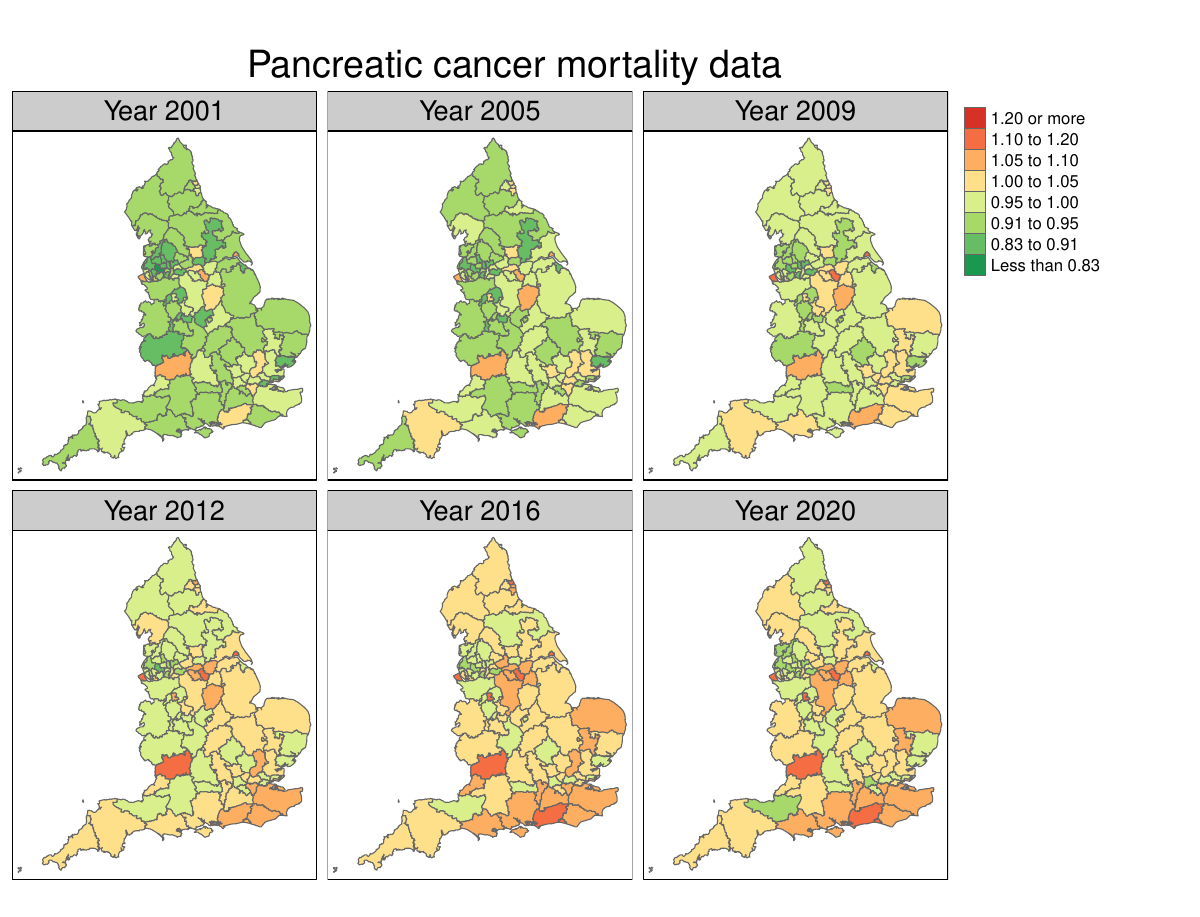}\\
        \vspace{-0.5cm}
        \includegraphics[width=0.9\textwidth]{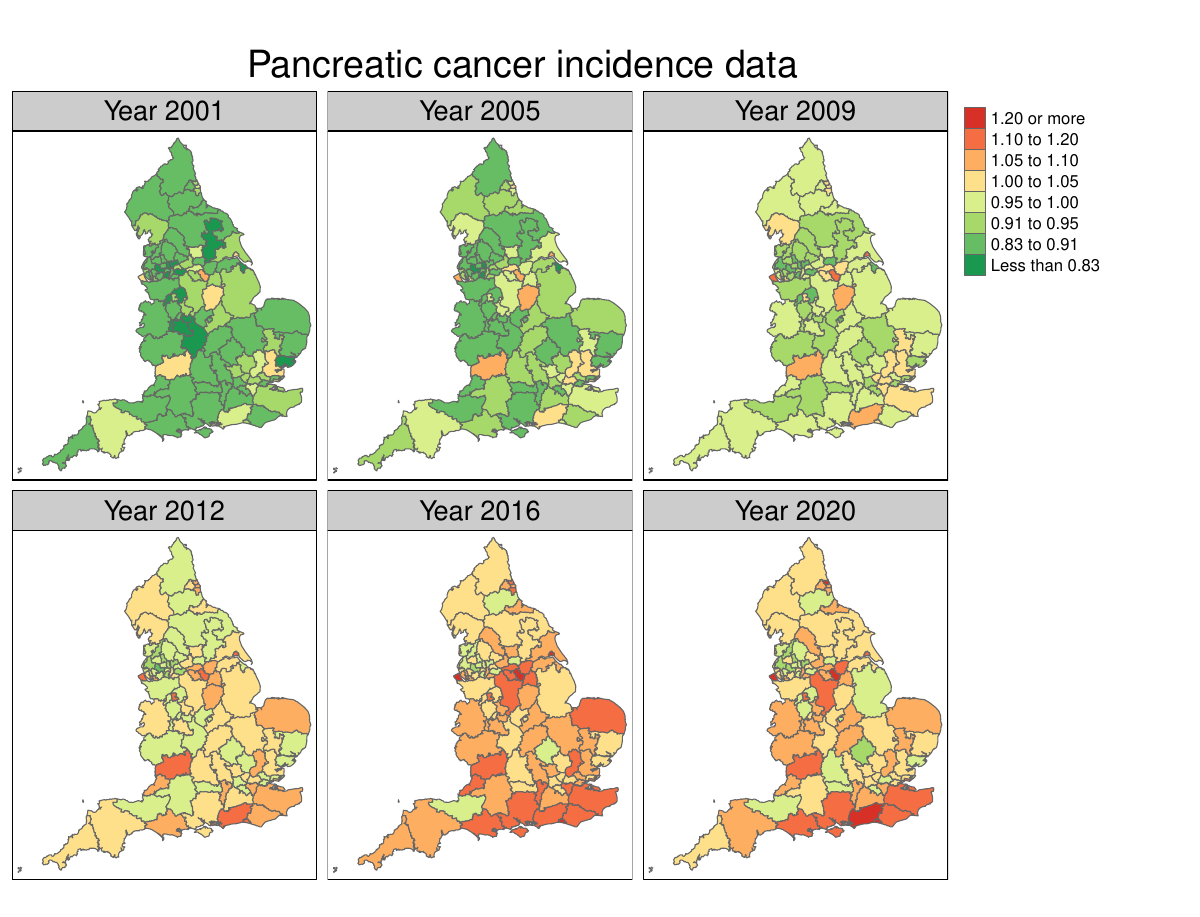}
        \vspace{-1cm}
    \end{center}
    \caption{Posterior median estimates of pancreatic cancer mortality (top) and incidence risks (bottom) for the shared-component model (M3) with Type II interaction effect. \label{fig:CancerData_Risks}}
\end{figure}

\autoref{fig:CancerData_Patterns} shows the maps with the posterior median estimates of the shared spatial random effects $\exp(\phi_i)$ (top left) and the posterior exceedence probabilities $P(\exp(\phi_i)>1 | \boldsymbol{y})$ (top right) obtained under the M3-TypeII model. Dark blue areas in the maps indicate significantly high area-specific risks during the period 2001-2020 in comparison to England as a whole. Posterior median estimates and 95\% credible intervals of the temporal shared random effect $\exp(\gamma_t)$ is also plotted (bottom). \autoref{fig:CancerData_Risks} display the maps with posterior median estimates of pancreatic cancer mortality (top) and incidence (bottom) risks.

\clearpage
\subsection*{Spatial Compositional Data: the case of \textit{Arabidopsis thaliana}}

\begin{figure}[!ht]
    \vspace{0.5cm}
    \includegraphics[width=\textwidth]{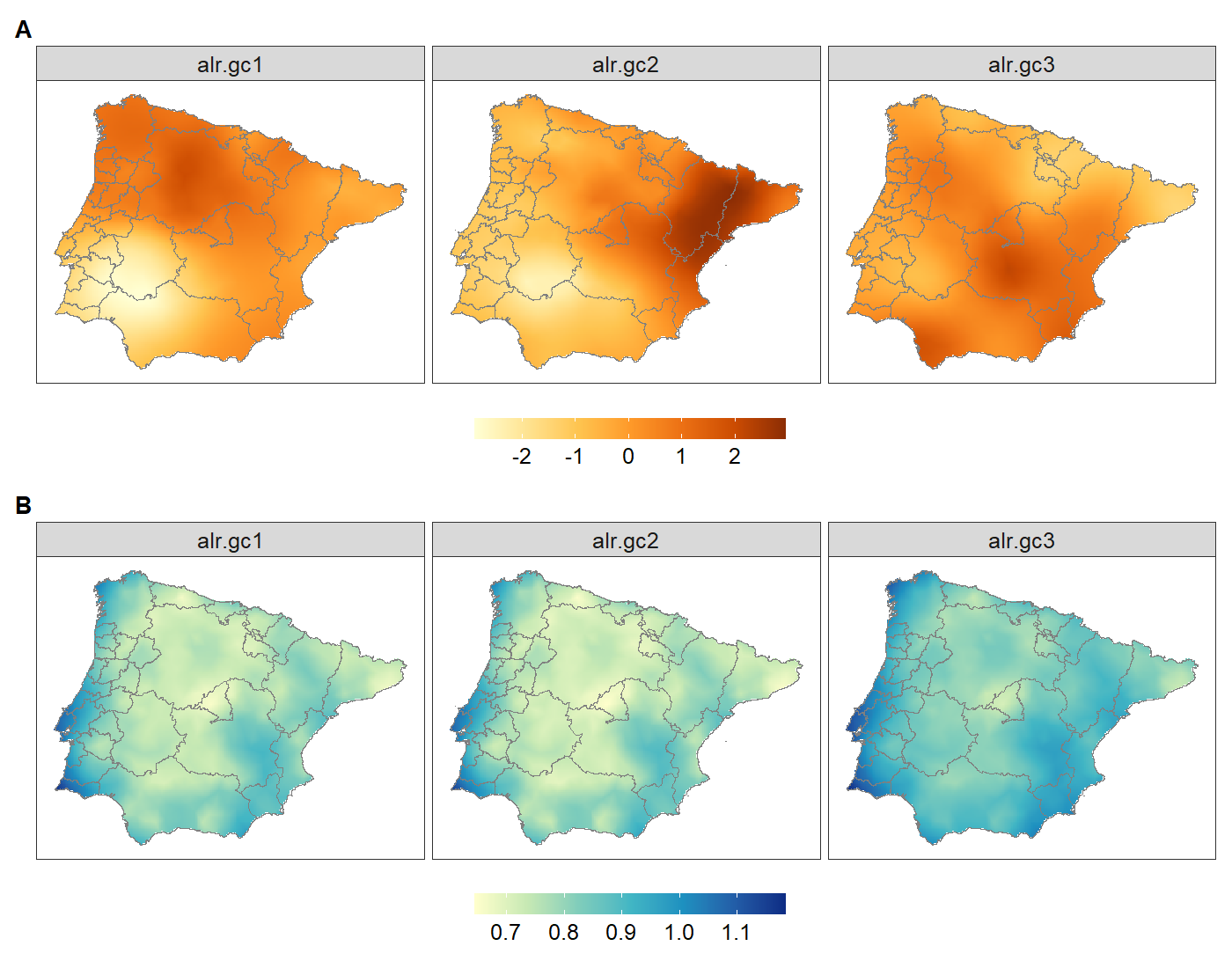}
    \vspace{-0.8cm}
    \caption{Posterior mean (\textbf{A}) and standard deviation (\textbf{B}) of the spatial effects by $alr$-coordinate under Type VIII structure model. \label{fig:coda_posterior_spatial}}
\end{figure}

\autoref{fig:coda_posterior_spatial} shows the posterior estimates (mean and standard deviation) of the spatial effects $\boldsymbol{\omega}^{(d)}, d=1,2,3$ for each $alr$-coordinate. We observe how the posterior spatial pattern shifts for each $alr$-coordinate, a crucial factor for obtaining more flexible predictions. 

\clearpage
\subsection*{Space-time models for the United Kingdom wind speed data}

\begin{figure}[!ht]
    \begin{center}
    \vspace{-0.5cm}
    \includegraphics[width=0.69\textwidth]{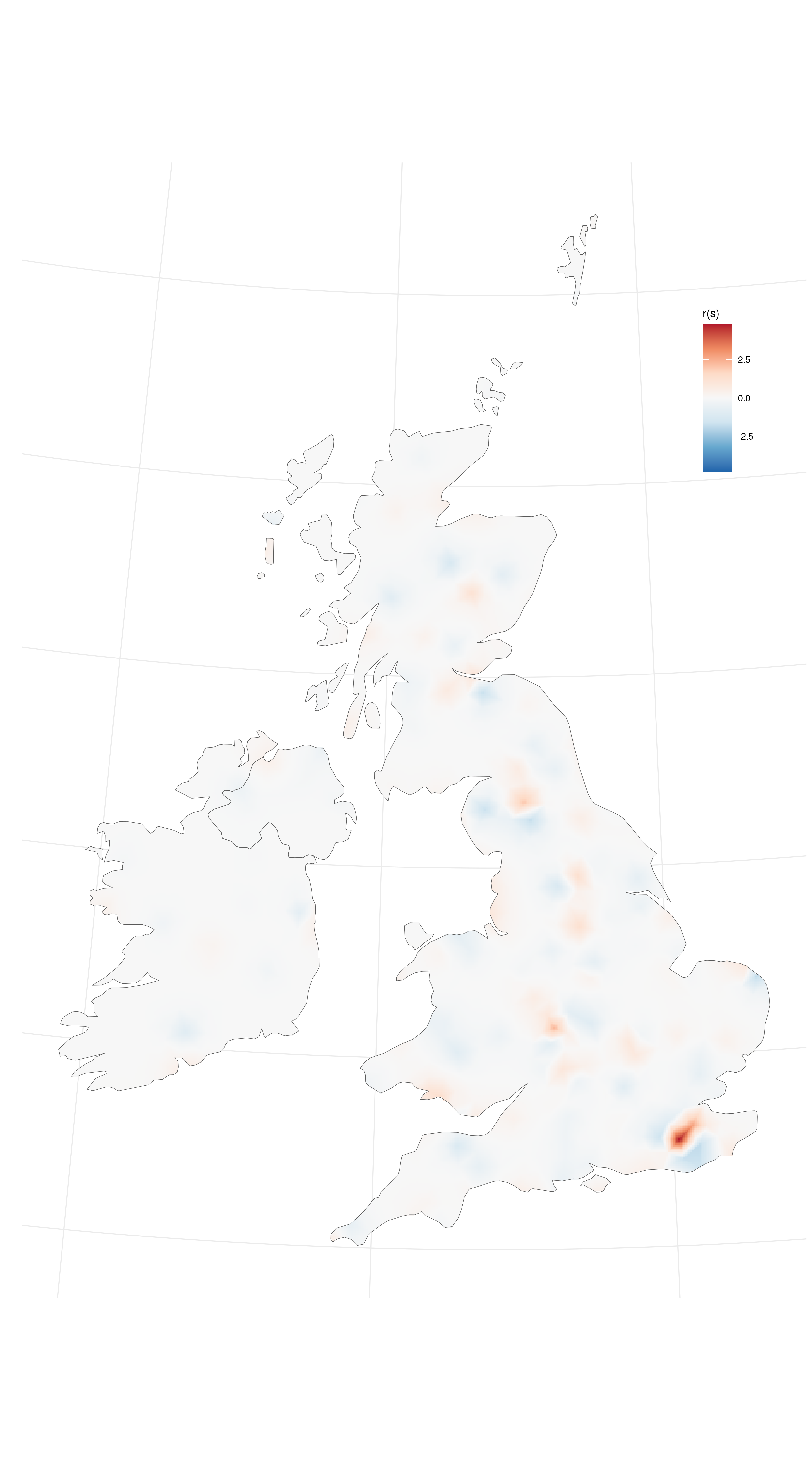}
    \vspace{-2cm}
    \end{center}
    \caption{Posterior mean of the spatial effect $r(\mathbf{s})$ from the fitted model B. \label{fig:Spacetime_u1mean}}
\end{figure}

\begin{figure}[!ht]
    \begin{center}
    \includegraphics[width=\textwidth]{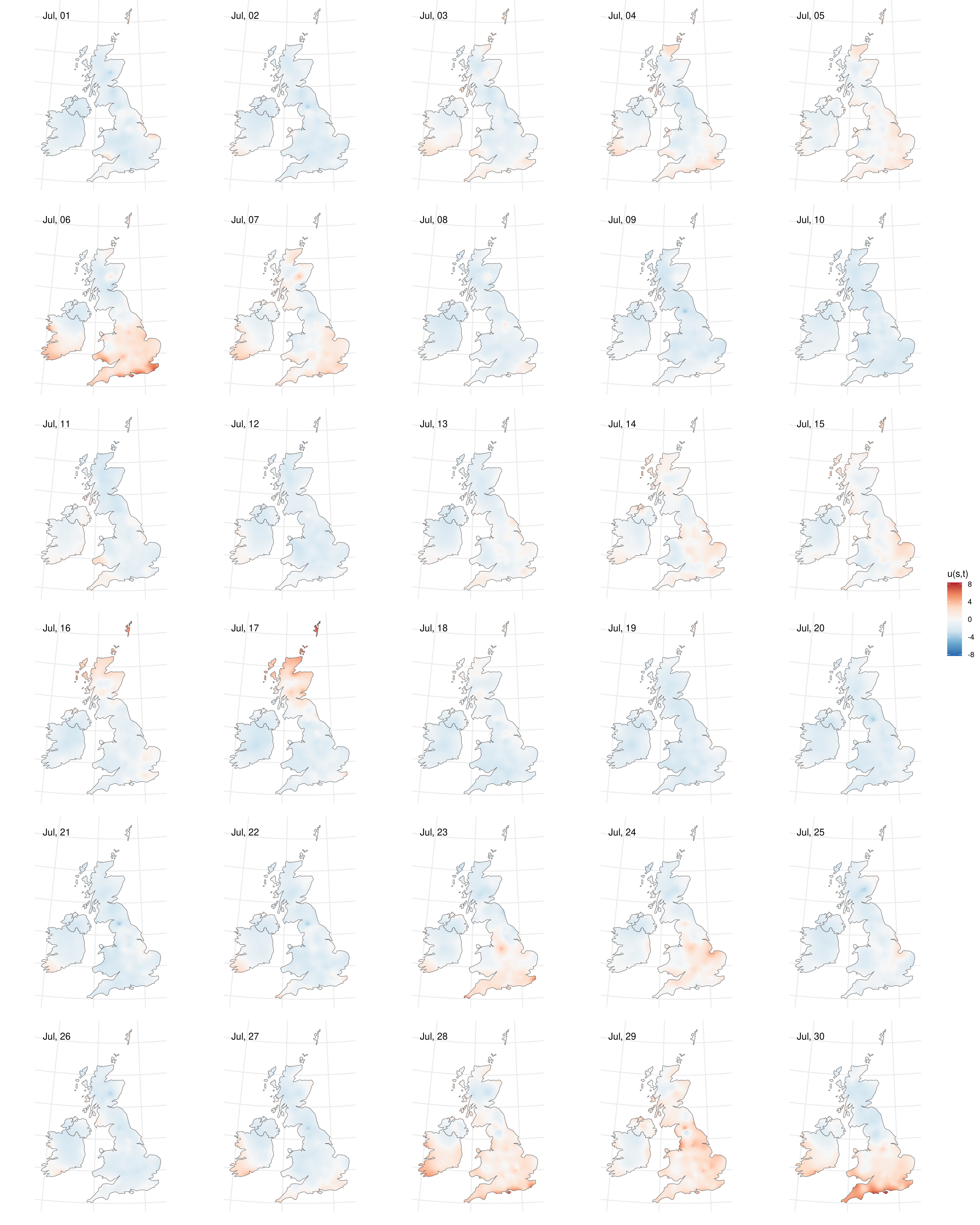}
    \vspace{-0.5cm}
    \end{center}
    \caption{Posterior mean of the space-time effect $u(\mathbf{s},t)$ from the fitted model B. \label{fig:Spacetime_u1mean}}
\end{figure}

\end{document}